\documentclass[twocolappendix]{emulateapj}

\usepackage{soul,color,amsmath}

\shorttitle{DISCO Code}
\shortauthors{Duffell}

\begin{document}

\author{Paul C.~Duffell}
\affil{Astronomy Department and Theoretical Astrophysics Center, University of California, Berkeley}
\email{duffell@berkeley.edu}

\title{DISCO: a 3D Moving-Mesh Magnetohydrodynamics Code Designed for the Study of Astrophysical Disks}

\begin{abstract}

This work presents the publicly available moving-mesh magnetohydrodynamics code DISCO.  DISCO is efficient and accurate at evolving orbital fluid motion in two and three dimensions, especially at high Mach number.  DISCO employs a moving-mesh approach utilizing a dynamic cylindrical mesh that can shear azimuthally to follow the orbital motion of the gas.  The moving mesh removes diffusive advection errors and allows for longer timesteps than a static grid.  Magnetohydrodynamics is implemented in DISCO using an HLLD Riemann solver and a novel constrained transport scheme which is compatible with the mesh motion.  DISCO is tested against a wide variety of problems, which are designed to test its stability, accuracy and scalability.  In addition, several magnetohydrodynamics tests are performed which demonstrate the accuracy and stability of the new constrained transport approach, including two tests of the magneto-rotational instability (MRI); one testing the linear growth rate and the other following the instability into the fully turbulent regime.

\end{abstract}


\section{Introduction}
\label{sec:intro}

The study of gaseous disks is of fundamental importance to astrophysics.  Disks are ubiquitous; essentially any gaseous orbital system which efficiently loses energy while conserving angular momentum will form a disk.  The formation of planets takes place in a protoplanetary disk, which can influence the first few million years of the planets' existence \citep[e.g.][]{2012ARAnA..50..211K, 2014prpl.conf..339T, 2014prpl.conf..475A}.  These especially include ``transition disks" whose cavities may be signposts of planet formation \citep[e.g.][]{2014prpl.conf..497E}.  Accretion disks around black holes (in particular, X-Ray binaries) are the most robust means of a black hole's detection via electromagnetic waves \citep[e.g.][]{2006ARAnA..44...49R, 2013LRR....16....1A}.  Some stars are also thought to be surrounded by an accreting disk \citep[e.g.][]{2013AnARv..21...69R}.  Accretion disks are also an important feature of cataclysmic variables \citep[e.g.][]{1976ARAnA..14..119R, 1978MNRAS.182..423P}, and they are thought to be the power source behind active galactic nuclei \citep[e.g.][]{1993ApJ...404..551O, 2000ARAnA..38..521S}.  Current efforts to observe the horizon of Sgr A* depend on an emitting disk of gas surrounding this supermassive black hole \citep[e.g.][]{2011JPhCS.283a2030P}.  Circumbinary disks are among the most promising possibilities for electromagnetic counterparts of gravitational wave emission from merging black holes \citep[e.g.][]{2013MNRAS.436.2997D, 2013ApJ...774..144R, 2014ApJ...783..134F, 2016PhRvL.116f1102A, 2016arXiv160204226S}.  Much of the same circumbinary physics applies to a newly-born binary star system, surrounded by a common protostellar disk \citep[e.g.][]{2007prpl.conf..395M}.  Galaxies can be considered another important type of disk \citep[e.g.][]{2013AnARv..21...61R, 2013pss6.book...91G}.  Even Saturn's rings constitute a disk, though it is not composed of gas, but of icy solids \citep[e.g.][]{1982ARAnA..20..249G}.

Each particular instance of a disk in nature possesses its own specific physical ingredients.  Many disks are ionized and therefore magnetic fields are important to their evolution.  Protoplanetary disks are dusty and also subject to the gravitational influence of the planets being formed in the disk.  Black hole accretion disks can constitute very extreme environments where radiation hydrodynamics, weak sector couplings, and general relativity all come into play.  In galaxies, self-gravity is very important, as opposed to many other systems which can be approximated as orbiting a single point mass at the center.  Nevertheless, many of the same techniques can be applied to study the physics behind this wide range of systems, since the most important ingredients (orbital and gas dynamics) are common to all of them.

Arguably the most convenient experimental test-beds for disk dynamics are numerical calculations.  The hydrodynamical equations governing gas dynamics and magnetohydrodynamics (MHD) have been integrated numerically using many approaches.  In astrophysics, the most commonly employed techniques are particle-based methods like smoothed particle hydrodynamics \citep[SPH,][]{1985AnA...149..135M, 2005MNRAS.364.1105S, 2009NewAR..53...78R, 2012JCoPh.231..759P}, and grid-based high-resolution shock-capturing techniques \citep{CPA:CPA3160100406, godunov1959difference, 1977JCoPh..23..276V, 1984JCoPh..54..115W, 1988JCoPh..75..400B, 2005JCoPh.205..509G, 2007ApJS..170..228M}, which use Godunov-type schemes for hydrodynamic evolution and often employ adaptive mesh refinement \citep[AMR,][]{2000ApJS..131..273F, 2002AnA...385..337T, 2004astro.ph..3044O} to resolve large dynamic ranges.  Recently, moving-mesh techniques \citep{2010MNRAS.401..791S, 2011ApJS..197...15D, 2015ApJS..216...35Y} and several new ``meshless" techniques \citep{2012ApJS..200....6M, 2015MNRAS.450...53H, 2015arXiv151200386D, 2016MNRAS.455...51H} have emerged as an attempt to merge the accuracy of the AMR approach with the flow adaptivity of SPH.  Moving mesh methods have already enjoyed remarkable success using an adaptive Voronoi tessellation for the shape of the mesh zones.  On the other hand, none of these approaches are specifically tailored to the special challenges inherent in disks.

Disks are often highly supersonic (Mach number $\mathcal{M} \gtrsim 10$).  Standard Godunov-type methods can stably evolve such supersonic flows, but this entails a diffusive upwind flux at every zone on every timestep.  This means one often needs high resolution for accuracy, and even at low resolution it places strong constraints on the allowed timestep, which is limited by the Courant condition.  It may be difficult to resolve complicated, potentially turbulent flows if the gas is being diffusively passed from cell to cell at high Mach number.

Several of the aforementioned techniques are designed to address this problem.  SPH and Voronoi codes are designed to adapt to supersonic flows by moving with the flow, effectively subtracting off the supersonic orbital motion.  On the other hand, many SPH formulations require significant artificial viscosity, and some do not converge at first order \citep[though this is not as devastating a problem for some modern SPH formulations, e.g.][]{2010MNRAS.405.1513R, 2015MNRAS.448.3628R}.  The Voronoi technique is designed to resolve such flaws, and in fact it has been suggested that the Voronoi mesh may be ideal for studies of disk-planet interactions \citep{2014MNRAS.445.3475M}, but the choice of Voronoi cells to describe the flow may be overkill, since the bulk orbital motion is very simple and is known before running the code.  There is also inherent noise generated when Voronoi cells shear past one another, as mesh faces can rotate rapidly \citep{2015MNRAS.449.2718D, 2015MNRAS.452.3853M, 2016MNRAS.tmp..569S}.  Because of these complications, the Voronoi moving-mesh technique may be better-suited to more general (e.g. cosmological) flows, where the large-scale bulk motion is not known at run-time.  In disks, angular momentum conservation is also of vital importance; a method which does not precisely maintain this circular orbital motion over thousands of orbits may struggle to capture small perturbations to this motion \citep{2015MNRAS.450...53H}.

Fortunately, because disks are so important, several codes have been written specifically to tackle disk dynamics.  The most well-known and well-tested is almost certainly FARGO \citep{2000AnAS..141..165M}.  FARGO is not just a code, but it is a general numerical technique, also called ``orbital advection", which has also been implemented in other grid-based codes \citep[e.g.][]{2008ApJS..177..373J, 2010ApJS..189..142S, 2012ApJ...749..189S}.  FARGO subtracts off the Keplerian component of the flow, by shifting zones some integral amount each timestep to account for the advective motion.  The residual advection with respect to the grid is accounted for by a standard Godunov flux.  A special version of FARGO designed for 3D disks has also recently been written \citep{2016arXiv160202359B}.

Another well-known code tailored to disks is the RODEO code \citep{2006AnA...450.1203P}.  RODEO solves the hydrodynamical equations in a (rigidly) rotating frame, and uses a special integral form of the field equations which is tailored for stability.  The large-scale orbital motion is subtracted, but there is still a nonzero shear flow with respect to the grid.  One of the advantages to RODEO is that mesh refinement is possible (this does not appear to be straightforward in the original FARGO scheme, since the orbital advection technique appears to place restrictions on the topology of the grid).  RODEO can therefore capture very accurate details in the vicinity of a planet, since its orbital motion is subtracted and the mesh can also be refined around the planet.

There is also an efficient code called PEnGUIn \citep{2014ApJ...782...88F}.  PEnGUIn is particularly efficient, as it has been optimized for use on graphics processors (GPUs).  This is ideal for studies of disks, as many disk-related problems require evolving the system over many thousands of orbits, which is often prohibitively expensive.  Large parameter surveys are seldom undertaken for these reasons, but PEnGUIn makes such expensive problems much more manageable.

\begin{figure}
\epsscale{1.0}
\plotone{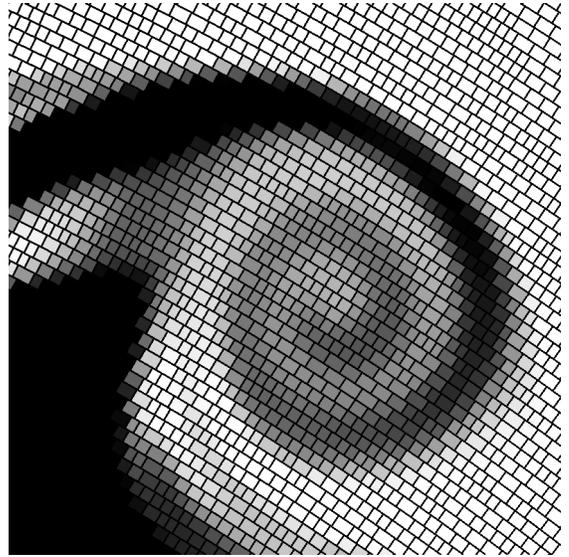}
\caption{ DISCO's numerical grid is shown in 2D, illustrating the rotating annular wedges which make up the cylindrical mesh.  The test problem being run here is the ``Cylindrical Kelvin Helmholtz" test of section \ref{sec:kh}.
\label{fig:gridimage} }
\end{figure}

While these existing methods are reliable and have demonstrated themselves useful for evolving disks in many scenarios, it is worthwhile to consider another distinct numerical approach.  In this work, the code DISCO is presented, a moving-mesh technique similar to the numerical scheme of the moving-mesh codes AREPO and TESS.  In the same spirit of FARGO and RODEO, orbital motion is subtracted, but unlike RODEO the entire shear flow can be subtracted, and unlike the original FARGO scheme, the topology of the numerical mesh is not restricted (this last point is particularly important if one does not wish to excise the inner disk).  The computational domain is decomposed into zones which are cylindrical wedges (Figure \ref{fig:gridimage}).  These zones are given an azimuthal velocity and are allowed to shear past one another smoothly.  This azimuthal velocity can be chosen to have any value, with the choice of the local fluid velocity resulting in an azimuthally Lagrangian scheme.  Because the mesh moves with the flow instead of passing the fluid from one zone to the next, advection errors resulting from orbital motion are significantly reduced and otherwise subtle features can be captured accurately while the flow orbits supersonically.

DISCO has already been applied to many challenging problems in astrophysics, including gap opening and orbital evolution in protoplanetary disks \citep{2012ApJ...755....7D, 2013ApJ...769...41D, 2014ApJ...792L..10D, 2015ApJ...806..182D, 2015ApJ...812...94D} and the evolution of circumbinary disks surrounding supermassive black hole binaries \citep{2014ApJ...783..134F, 2015MNRAS.446L..36F, 2015MNRAS.447L..80F, 2016MNRAS.tmp..577D}.  In this work, the numerical technique is described in detail (Section \ref{sec:numerics}).  This includes the recent addition of a constrained transport method for solving the equations of magnetohydrodynamics (Section \ref{sec:mhd}).  In Section \ref{sec:tests}, a series of numerical code tests is presented, to demonstrate the convergence and practicality of the code.  Results are summarized in Section \ref{sec:discussion}.

\section{Numerical Method}
\label{sec:numerics}

\subsection{Field Equations}

DISCO is capable of evolving arbitrary hyperbolic partial differential equations in conservation-law form.  Its simplest mode solves Euler's equations:

\begin{equation}
\begin{array}{c}
\displaystyle \partial_t ( \rho ) + \nabla \cdot ( \rho \vec v ) = 0 \\
\displaystyle \partial_t ( \rho \vec v ) + \nabla \cdot ( \rho \vec v \vec v + P \tensor{I} ) = 0 \\
\displaystyle \partial_t ( \frac12 \rho v^2 + \epsilon ) + \nabla \cdot ( ( \frac12 \rho v^2 + \epsilon + P ) \vec v ) = 0 
\end{array} 
\label{eqn:bare}
\end{equation}
where $\rho$ is density, $\vec v$ is velocity, $P$ is pressure, and $\epsilon$ is internal energy density.  An adiabatic equation of state is typically employed:

\begin{equation}
P = (\gamma - 1)\epsilon,
\end{equation}
where $\gamma$ is the adiabatic index.  Additional terms such as viscosity, gravity, and magnetic fields will be included in later subsections.  For now, the numerical formulation will be expressed in terms of these ``bare" equations.  Other forms, such as the special and general relativistic versions of these equations, will not be discussed here, but will be addressed in a future work.

Because angular momentum conservation is so important to the orbital dynamics, the momentum conservation law is evaluated in terms of the vertical, radial, and angular momentum:

\begin{equation}
\begin{array}{c}
\displaystyle \partial_t ( r^2 \rho \omega ) + \nabla \cdot ( r^2 \rho \omega \vec v + P \hat \phi ) = 0 \\
\displaystyle \partial_t ( \rho v_r ) + \nabla \cdot ( \rho v_r \vec v + P \hat r ) = \rho \omega^2 r + P/r \\
\displaystyle \partial_t ( \rho v_z ) + \nabla \cdot ( \rho v_z \vec v + P \hat z ) = 0
\end{array} 
\label{eqn:angmom}
\end{equation}
where $r$ is the cylindrical radius, $\omega$ is the angular fluid velocity, and $v_r$ and $v_z$ are the radial and vertical components of velocity.  The source terms on the right-hand side of the equation for radial momentum come from the evaluation of the tensor divergence in cylindrical coordinates.  This reflects the fact that ``radial momentum" is not a conserved quantity.  These terms can be interpreted as a centrifugal force, and an azimuthal pressure-balance term that comes out of the non-zero divergence of the $\hat r$ vector.

Finally, before discretizing these equations, the energy equation is modified to improve accuracy.  Define the quantity 

\begin{equation}
\tilde \omega \equiv \omega - \Omega_E(r), 
\end{equation}
where $\Omega_E(r)$ is some differentiable function of radius.  It should be made clear that $\Omega_E(r)$ has nothing to do with the mesh motion described in the later sections; it is defined in order to subtract off part of the kinetic energy from the field equations.  For typical applications, $\Omega_E(r)$ will be the Keplerian orbital velocity, but in principle it can be any analytically known differentiable function of radius, including $\Omega_E = 0$.  Nothing in the DISCO algorithm depends sensitively on the choice of $\Omega_E(r)$, but in practice subtracting off this large kinetic component of the energy can yield a substantial improvement to the stability and reliability of the numerical scheme, especially for high Mach number flows.  Also, note that $\Omega_E(r)$ will not be subtracted from the angular momentum, so that Coriolis terms do not appear in the momentum equations.

Subtracting $\Omega_E(r)$ from $\omega$ in the energy equation yields the following evolution equation (where $\tilde v$ is the velocity with $\Omega_E$ subtracted):

\begin{eqnarray}
&& \partial_t ( \frac12 \rho \tilde v^2 + \epsilon )  \nonumber \\
&+& \nabla \cdot ( ( \frac12 \rho \tilde v^2 + \epsilon + P ) \vec v )  \nonumber \\
&=& r \rho v_r ( \Omega^2_E(r) - r {d\Omega_E \over dr} \tilde \omega ) 
\end{eqnarray}

If $\Omega_E(r) \neq 0$ is chosen, energy is not explicitly conserved in this formulation, due to the presence of this source term.

To summarize, the field equations can be expressed in the conservation-law form:

\begin{equation}
\partial_t u + \nabla \cdot \vec F = S.
\label{eqn:conslaw}
\end{equation}

\begin{figure}
\epsscale{1.0}
\plotone{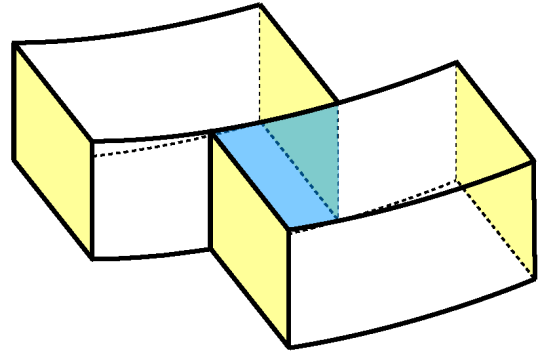}
\caption{ Diagram of two adjacent cells in DISCO's mesh.  The blue shaded region represents a ``face" in DISCO; it is defined as the region of overlap between the surfaces of two neighboring zones.
\label{fig:faceimage} }
\end{figure}

The conservation laws can be expressed in terms of the five primitive variables:
\begin{equation}
W = \{ \rho , \omega , v_r , v_z , P \}.
\end{equation}
Now, the evolution equations can be compactly summarized by writing down the five conserved variables,
\begin{equation}
u = \{ \rho , r^2 \rho \omega , \rho v_r , \rho v_z , \frac12 \rho \tilde v^2 + \epsilon \},
\label{eqn:bareu}
\end{equation}
the five corresponding fluxes,
\begin{eqnarray}
\vec F &=& \{ ~ \rho \vec v ~, ~r^2 \rho \omega \vec v + P \hat \phi~ , ~\rho v_r \vec v + P \hat r~ , \nonumber \\
&&  ~\rho v_z \vec v + P \hat z~ , ~( \frac12 \rho \tilde v^2 + \epsilon + P ) \vec v~ \}, 
\end{eqnarray}
and the five source terms,
\begin{eqnarray}
S &=& \{ 0 , 0 , \rho \omega^2 r + P/r , 0 , \nonumber \\
&& r \rho v_r ( \Omega^2_E(r) - r \Omega'_E(r) \tilde \omega ) \}.
\end{eqnarray}

Now that these have been specified, the evolution of the system can be described in terms of the generic expression (\ref{eqn:conslaw}).

\subsection{Mesh Construction Algorithm}
\label{sec:mesh}

Equation (\ref{eqn:conslaw}) will be discretized in the following subsection (\ref{sec:integral}).  First, it is necessary to describe DISCO's mesh, and how it is constructed.

Similar to the formulation of the TESS code \citep{2011ApJS..197...15D}, the numerical scheme can be completely specified, given the volumes of the zones, logical information specifying which zones are neighbors, and the areas of the ``faces" connecting neighboring zones.  The zones are annular wedges with extents given in cylindrical coordinates by $\Delta r$, $\Delta \phi$ and $\Delta z$ (Figure \ref{fig:faceimage}). The faces with $\hat \phi$ normal are the ``front" and ``back" of these zones (yellow shaded area of Figure \ref{fig:faceimage}).  The faces with $\hat r$ or $\hat z$ normal are defined as the overlap of the boundary of two neighboring zones (blue shaded area of Figure \ref{fig:faceimage}).  This means that zones can have more than two radial or vertical faces (on average, zones typically have four of each).

Given neighboring annuli at radii $r_j$ and $r_{j+1}$, first the zone at each of these radii intersecting the radial ray $\phi = 0$ are found.  These two zones are guaranteed to share a face.  The geometry of the shared face is identified, and then the next face is found by advancing whichever of the two zones has a smaller $\phi_{i+1/2}$ associated with their front face.  Again, this new pair of zones must share a face.  This procedure is repeated for $N_{\phi}(j) + N_{\phi}(j+1)$ faces, where $N_{\phi}(j)$ is the number of zones in a given annulus labeled by the index $j$.  Each step, a face is identified as the intersection of the boundary of the pair of zones.

Faces normal to $\hat z$ are constructed in an analogous way.

\subsection{Integral Form / Mesh Motion}
\label{sec:integral}

DISCO is a finite-volume method.  To discretize the system, (\ref{eqn:conslaw}) is integrated over the volume of a computational zone, using Gauss' law on the flux term:

\begin{equation}
\int \partial_t u dV + \oint \vec dA \cdot \vec F = \int S dV.
\label{eqn:intcons}
\end{equation}

Now, at first, consider the case of no mesh motion, for which 

\begin{equation}
\int ( \partial_t u ) dV = \partial_t \int u dV,
\label{eqn:fixedzone}
\end{equation}
and define $M^n_i$ to be the amount of a given conserved quantity in zone i at timestep n:

\begin{equation}
M^n_i \equiv \int u dV
\end{equation}

After performing an integral in time, it is then straightforward to write an evolution equation for the $\{M^n_i\}$:

\begin{equation}
M^{n+1}_i = M^n_i - \Delta t \sum\limits_{\text{face}~f} \vec {dA}_{f} \cdot \vec F_{f} - S_i \Delta t \Delta V.
\label{eqn:evol1}
\end{equation}
where the sum is over faces bounding zone $i$.  So far, no approximations have been made, so long as $\vec F_{f}$ is interpreted as the time-averaged and face-averaged flux, and $S_i$ is interpreted as the volume-averaged and time-averaged source term.  Exact geometry is employed, so that for example $\Delta V$ is the exact volume of a cylindrical wedge:

\begin{equation}
\Delta V = \Delta \phi \Delta z ( \frac12 r_{+}^2 - \frac12 r_{-}^2 ),
\end{equation}
where $\Delta \phi$ and $\Delta z$ are the azimuthal and vertical extent of the zone, and $r_{+}$ and $r_{-}$ are the outer and inner radii.  Some of the conserved quantities and source terms require the coordinate r.  In this case, the radius is chosen to be the moment arm of the zone:

\begin{equation}
r_{\rm moment} = \sqrt{ \frac12 ( r_{+}^2 + r_{-}^2 ) }
\end{equation}

If the orbital motion of the grid is switched on, the integral form of these equations departs from (\ref{eqn:evol1}), since the control volumes and their associated faces move through space.  In this case, the generalization of Equation (\ref{eqn:fixedzone}) is given by the Reynolds transport theorem as

\begin{equation}
\int (\partial_t u) dV = \partial_t \left( \int u dV \right) - \oint u \vec w \cdot \vec {dA},
\end{equation}
where $\vec w$ is the velocity of the boundary of the zone.  This results in the following modification to Equation (\ref{eqn:evol1}):

\begin{equation}
M^{n+1}_i = M^n_i - \Delta t \sum\limits_{\text{face}~f} \vec {dA}_{f} \cdot ( \vec F_{f} - \vec w_{f} u_{f} ) - S_i \Delta t \Delta V.
\label{eqn:evolution}
\end{equation}

In other words, the time evolution differs from a standard Godunov-type method by making the substitution $\vec F \rightarrow \vec F - \vec w u $, where $\vec w$ is the velocity of the face upon which the flux is evaluated, and now $u$ on the face needs to be determined in addition to $F$.  For DISCO's cylindrical mesh, the velocity $\vec w$ is always normal to the face.  The face velocities are zero for all radially and vertically oriented faces; all mesh motion is azimuthal.  The azimuthal velocity can be chosen in many ways.  If desired, each azimuthal face can be moved independently, according to the average velocity of its adjacent zones.  Alternatively, some global analytical formula for $\vec w(r)$ can be prescribed, if one does not wish for the zones to have too much independence.  The mesh velocity could also be set to $w = r \Omega_E(r)$, where $\Omega_E(r)$ is the analytical function described in the previous subsection.  Again, it is not necessary to force these two functions to be equal.

So far, no numerical approximations have been made; equation (\ref{eqn:evolution}) is merely an integral form of (\ref{eqn:conslaw}).  Therefore, the numerical approximations are housed in the estimation of $\vec F_{f}$, the time-averaged and area-averaged flux through the face (as well as $u_{f}$ and $S_i$).  These numerical approximations are detailed in the following two subsections.

\subsection{Riemann Solver}

\begin{figure}
\epsscale{1.0}
\plotone{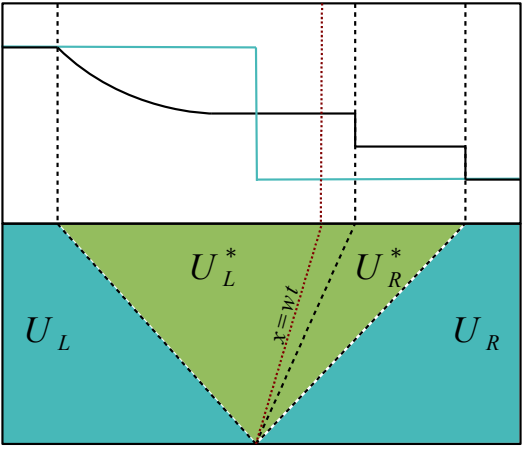}
\caption{ Schematic diagram of the Riemann problem on a moving mesh.  The face velocity is traced out by the red dashed curve $x=wt$; this is the characteristic on which the solution is evaluated.
\label{fig:riemann} }
\end{figure}

Equation (\ref{eqn:evolution}) requires a numerical estimate for $F_{f}$, the time-averaged and area-averaged flux through face $f$.  The standard method for calculating such a flux in the Godunov method is to use a Riemann solver.

A Riemann solver takes as input a left and right state $\{u_L\}$, $\{u_R\}$ and returns as output some estimate of the solution to the shock-tube problem given by piecewise-constant initial conditions:

\begin{equation}
u(x,t=0) = \left\{ \begin{array}{rl}
 u_L & ~ x < 0 \\
 u_R & ~ x > 0
       \end{array} \right.
\end{equation}

A standard Riemann solver takes this initial data and either computes the exact solution at a future time, or approximates it, returning the flux through the interface at $x=0$:

\begin{equation}
F_* = F(x=0,t)
\end{equation}

In the moving-mesh case, a different output is desired.  If the interface moves with velocity $w$, then the flux should be evaluated along the characteristic $x = wt$:
\begin{eqnarray}
F_* = F(x=wt,t) \\
u_* = u(x=wt,t)
\end{eqnarray}
(see Figure \ref{fig:riemann}).  Evaluating the Riemann solution along a given characteristic ($x=wt$ instead of $x=0$) is straightforward.

DISCO employs several different approximate Riemann solvers.  HLLE and HLLC \citep{toro2013riemann} are available for all flows, though HLLC is necessary to preserve contact discontinuities to high precision \citep{2011ApJS..197...15D}.  For MHD flows, an HLLD solver is implemented in DISCO, based on the solver of \cite{2005JCoPh.208..315M}.  Analogous to HLLC, the HLLD solver is necessary for preserving magnetic discontinuities to high precision (advection of a field loop can be solved to machine precision, but only if the HLLD solver is employed).


\subsection{Piecewise Linear Reconstruction}
\label{sec:plm}

\begin{figure}
\epsscale{1.0}
\plotone{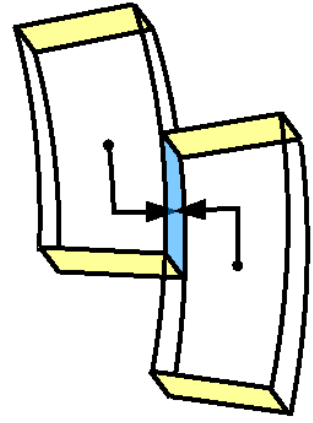}
\caption{ Extrapolation to the face adjoining two cells.  The misalignment of the cells means that the gradients in the azimuthal dimension are needed to extrapolate primitive variables to radially and vertically oriented faces.
\label{fig:plmimage} }
\end{figure}

In order to achieve second-order accuracy in space, primitive variables must be extrapolated from zone centers to faces to produce input to the Riemann solver (Figure \ref{fig:plmimage}):

\begin{eqnarray}
W_L = W_i + (\vec x_f - \vec x_i) \cdot (\vec \nabla W)_i \\
W_R = W_j + (\vec x_f - \vec x_j) \cdot (\vec \nabla W)_j
\end{eqnarray}
where the $\vec \nabla W$ are slopes which are estimated at the zone centers.  In calculating these slopes, care must be taken as the primitive variables cannot be assumed to represent differentiable functions.  Therefore, after estimating gradients of the primitive variables, a slope-limiter is applied to ensure stability in non-smooth regions of the flow.

First, the azimuthal gradients are calculated.  This is done by calculating left, right, and centered gradients in the zone:

\begin{eqnarray}
S_L &=& (W^i - W^{i-1})/( r \Delta \phi_L ) \\
S_R &=& (W^{i+1} - W^i)/( r \Delta \phi_R ) \\
S_C &=& (W^{i+1} - W^{i-1})/( r (\Delta \phi_L + \Delta \phi_R) ) 
\end{eqnarray}
where $\Delta \phi_L = (\Delta \phi_i + \Delta \phi_{i-1})/2$, and $\Delta \phi_R = (\Delta \phi_i + \Delta \phi_{i+1})/2$.  The slope-limited azimuthal gradient is then given by

\begin{equation}
\nabla_\phi W = \text{minmod}( \theta_{\rm plm} S_L , \theta_{\rm plm} S_R , S_C ),
\end{equation}

where $\theta_{\rm plm}$ is a slope-limiting parameter $1 < \theta_{\rm plm} < 2$ and the ``minmod" function is given by

\begin{equation}
\text{minmod}(x,y,z) = \left\{ \begin{array}{cl}
 \text{min}(x,y,z) & x,y,z~ > 0 \\
 \text{max}(x,y,z) & x,y,z~ < 0 \\
 0 & \text{otherwise}
       \end{array} \right.
\end{equation}

Next, the radial and vertical gradients are calculated.  For brevity, the formulas for the vertical gradients are omitted, as they are similar to the formulas for the radial gradients.

First, the radial gradient at each radially-oriented face is estimated, using the extrapolated values from the azimuthal gradient:

\begin{equation}
W_{if} \equiv W_i + r_i \Delta \phi_{if} (\nabla_\phi W)_i
\label{eqn:grad1}
\end{equation}

\begin{equation}
\left< \nabla_r W \right>^{\rm face}_{ij} = { W_{if} - W_{jf} \over r_i - r_j},
\end{equation}
where $\Delta \phi_{if}$ is the angular separation between the center of zone $i$ and the center of face $f$.  The zone-centered gradient is then estimated by performing an average over faces, weighted by face area:

\begin{equation}
\left< \nabla_r W \right>^{\rm zone}_{i} = {  \sum\limits_{j} dA_{j} \left< \nabla_r W \right>^{\rm face}_{ij} \over \sum\limits_{j} dA_{j}  }
\end{equation}

This provides a radial gradient in each zone, which like the azimuthal gradient must be processed through a slope limiter for stability.  This slope limiter is essentially the same as the one used in the azimuthal direction, but including all neighbors.  The ``centered" slope has already been calculated above; the final slope used is given by

\begin{equation}
\nabla_r W = \text{minmod}( \left< \nabla_r W \right>^{\rm zone}_{i} , \left\{ \theta_{\rm plm} \left< \nabla_r W \right>^{\rm face}_{ij} \right\} ).
\label{eqn:grad2}
\end{equation}

Formulas (\ref{eqn:grad1})-(\ref{eqn:grad2}) are then repeated for vertical gradients so that the gradient $\vec \nabla W$ is fully determined in the zone.  For most problems, the slope-limiting parameter is chosen to be $\theta_{\rm plm} = 1.5$.

\subsection{Time Evolution}

Equation (\ref{eqn:evolution}) specifies how to advance from timestep $n$ to timestep $n+1$, given the time-averaged values of $F$ and $u$ on the face.  This is given by the Riemann solver, which takes as input a left and right state, $\{W_L\}$ and $\{W_R\}$.  This left and right state are found by extrapolating from zone centers to face centers, using the slope-limited gradients given in section \ref{sec:plm}.

At this point, the system is completely specified, but the time-evolution operator, as expressed in equation (\ref{eqn:evolution}), is only first-order in time.  It has the following form:

\begin{equation}
M^{n+1}_i = M^{n}_i + \Delta t L_i( \{ \text{state} ~n \} ),
\end{equation}
where $L$ is a time-evolution operator depending on the state of the system at timestep $n$.  To increase the order of accuracy of the code, a method-of-lines technique is employed, introducing the intermediate state $M^{(1)}$:

\begin{eqnarray}
M^{(1)}_i &=& M^{n}_i + \Delta t L_i( \{ \text{state} ~n \} ), \\
M^{n+1}_i &=& \frac12( M^{(1)}_i + M^{n}_i ) + \frac12 \Delta t L_i( \{ \text{state} ~(1) \} ).
\end{eqnarray}

This constitutes a second-order timestep which is consistent with a total variation diminishing scheme.  The timestep $\Delta t$ is Courant-limited; that is,

\begin{equation}
\Delta t < \text{min}( \Delta t^{\rm cross}_i ),
\label{eqn:courant}
\end{equation}
where $\Delta t^{\rm cross}$ is the shortest signal-crossing time of a zone:

\begin{equation}
\Delta t^{\rm cross} = \text{min}\left( {\Delta r \over c_s + |v_r|}, {\Delta z \over c_s + |v_z|}, {r \Delta \phi \over c_s + |v_\phi - w|} \right).
\end{equation}

Note that moving the zones to cancel a supersonic orbital velocity ($w \sim v_\phi \gg c_s$) can increase the allowed timestep by orders of magnitude.  In the MHD case the sound speed in the above formula is replaced by the speed of fast magnetosonic waves.

The courant condition (\ref{eqn:courant}) is satisfied at each time-step by setting $\Delta t = C_{\rm CFL} \Delta t^{\rm cross}$, where $C_{\rm CFL} < 1$ (typically $0.5$ for hydrodynamical flows and $0.2$ for MHD).

\subsection{Parallelization}

\begin{figure}
\epsscale{1.0}
\plotone{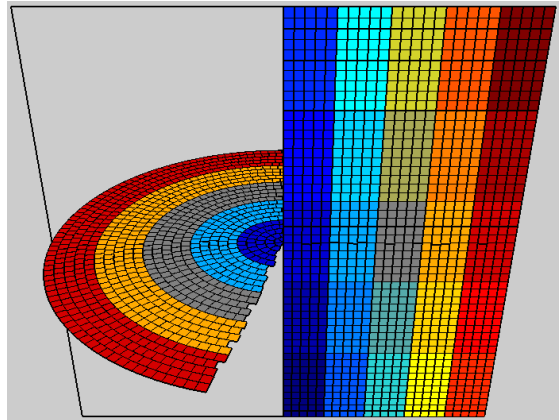}
\caption{ Parallelization is accomplished by domain decomposition in the radial and vertical dimensions.  Two cut planes are displayed in 3D showing which processor each zone belongs to.  Each color represents a single processor (in this example, the work is divided among 25 processors).
\label{fig:parallel} }
\end{figure}

DISCO achieves efficient parallelization by subdividing the computational domain into annuli.  Define $N^G_r$ and $N^G_z$ to be the global radial and vertical dimensions of the computational grid.  Define the indices $n_r$ and $n_z$ which label the radial and vertical grid: $0 < n_r < N^G_r$ and $0 < n_z < N^G_z$.  The number of zones in the azimuthal dimension can vary with $r$ and $z$: $N_\phi = N_\phi( n_r , n_z )$.  Typically $N_\phi$ is chosen at each radius so that the zones have a nearly $1:1$ aspect ratio.  The domain is then subdivided in $r$ and $z$ so that each processor has a local resolution $N^L_r$, $N^L_z$ (Figure \ref{fig:parallel}).  Boundary data is shipped in the vertical and radial direction every timestep.

It is also possible to subdivide the domain in azimuth, but such a subdivision adds significant complexity to the method.  The official version of DISCO therefore only performs parallel subdivision of the domain radially and vertically (in other words, $N^G_\phi = N^L_\phi = N_\phi$).  Performance and scaling of DISCO on thousands of CPUs is tested in section \ref{sec:scaling}.

\subsection{Disk-Satellite Interactions}

Disks usually orbit around a central point mass, meaning gravitational source terms must enter the evolution equations.  Additionally, there may be orbiting satellites exerting a gravitational influence on the disk.  These influences are accounted for by adding the following source terms to the energy and momentum equations:

\begin{equation}
S^{\rm Momentum}_{\rm grav} = \rho \vec g,
\end{equation}
\begin{equation}
S^{\rm Energy}_{\rm grav} = \rho ( \vec g \cdot \vec v )
\end{equation}
where $\vec g$ is the gravitational acceleration due to all point masses, labeled by $p$:

\begin{equation}
\vec g = \sum\limits_{p} - \nabla \Phi_p
\end{equation}
\begin{equation}
\Phi_p = -{G m_p \over ( |\vec x - \vec x_p|^n + \epsilon_p^n )^{1/n}}
\end{equation}

Here, $n$ and $\epsilon_p$ are optional smoothing parameters.  Typically in 2D calculations, $n=2$ and $\epsilon_p = 0.5 h$, where $h$ is a scale height.  This is to mimic the averaging of the gravitational force over the disk scale height (if there is a central body, typically $\epsilon_p = 0$ for that body).  In the code, all point masses (e.g. stars, planets, black holes) are simply called ``planets", and treated identically in an algorithmic sense.  The point masses can be given any prescribed motion $\vec x_p(t)$, or be moved due to the gravitational influence of the gas.

Accretion onto these bodies is also possible using an additional source term in the continuity equation.  This has been employed in studies of binary-disk interactions \citep{2014ApJ...783..134F, 2015MNRAS.446L..36F, 2015MNRAS.447L..80F, 2015arXiv151205788D}, but is not in DISCO's public version as there are many possible choices for such a term.

\subsection{Viscosity}
\label{sec:visc_sec}

A Navier-Stokes viscosity can be represented as a source term in the momentum equation,

\begin{equation}
\vec S = \nabla \cdot \tensor{\sigma},
\end{equation}
and a source term in the energy equation,

\begin{equation}
S = ( \nabla \cdot \tensor{\sigma} ) \cdot \tilde v + \sigma_{ij} \nabla_i v_j,
\end{equation}
where the viscous stress tensor $\tensor{\sigma}$ will be defined below.  The first term in the energy equation can be interpreted as work done by viscous forces (inner product of force with velocity, $F \cdot v$) and the second term expresses viscous heating.  Both of these source terms can be re-expressed as a viscous flux:

\begin{equation}
F^{\rm Momentum}_{\rm visc} = - \tensor{\sigma},
\end{equation}
\begin{equation}
F^{\rm Energy}_{\rm visc} = - \tensor{\sigma} \cdot \tilde v,
\end{equation}

This is possible because viscosity is an internal body-force in the gas, and therefore  conserves total momentum and energy.  In the case that $\Omega'_E(r) \ne 0$, the energy equation has a source term:

\begin{equation}
S^{Energy}_{\rm visc} = \sigma_{r \phi} r \Omega'_E(r).
\end{equation}

In cylindrical coordinates, the tensor divergence generates a more complicated expression for the fluxes, including a source term for radial momentum.  The viscous stress tensor $\tensor{\sigma}$ is proportional to the velocity gradients:

\begin{equation}
\sigma_{ij} = \nu \rho ( ( \nabla_i v_j + \nabla_j v_i ) + \eta \delta_{ij} \nabla \cdot v )
\end{equation}

$\eta$ is a dimensionless order-unity constant which summarizes the relationship between bulk and shear viscosity.  For most orbital flows, the choice of $\eta$ is unimportant (in fact, for most problems it will only be the value of $\sigma_{r \phi}$ that matters).  The full derivation appears in the appendix; the final values of the viscous flux and source terms are

\begin{eqnarray}
\vec F_{\rm visc} = -\nu \rho \{ 0 , r^2 \vec \nabla \omega + 2 v_r \hat \phi , \vec \nabla v_r - 2 \omega \hat \phi , \vec \nabla v_z , \nonumber \\
v_r \vec \nabla v_r + r^2 \tilde \omega \vec \nabla \omega + v_z \vec \nabla v_z - 2 r \omega \tilde \omega \},
\label{eqn:viscflux}
\end{eqnarray}

\begin{equation}
S_{\rm visc} = \{0,0,-\nu \rho v_r / r^2, 0, -\rho \nu (\nabla_{\phi} v_r + r \nabla_r \omega) r \Omega_E'(r) \}.
\label{eqn:viscsrc}
\end{equation}
where $\tilde \omega = \omega - \Omega_E(r)$.  In section \ref{sec:visctest}, several test problems will be presented to empirically check that all of these terms are correct.

Note that the formula (\ref{eqn:viscflux}) is expressed in terms of gradients of the primitive variables.  The viscous flux is evaluated separately from the Riemann solver.  The primitive variables $W$ are extrapolated to the face to attain the quantities $W_L$ and $W_R$, and these quantities and their slope-limited gradients $\vec \nabla W$ are averaged between the left and right state, before using them to evaluate the viscous fluxes (\ref{eqn:viscflux}).

\subsection{Magnetohydrodynamics}
\label{sec:mhd}

\begin{figure}
\epsscale{1.0}
\plotone{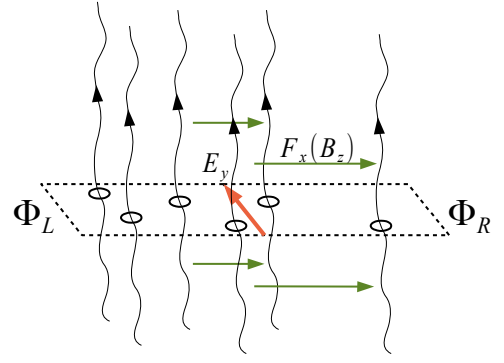}
\caption{ Schematic diagram of Faraday's law in integral form; the flow of magnetic field lines across an edge is equal to the line integrated electric field along the edge.
\label{fig:eflux1} }
\end{figure}

Similar to Euler's equations, the field equations of MHD can also be expressed in conservation-law form:

\begin{equation}
\begin{array}{c}
\displaystyle \partial_t ( \rho ) + \nabla \cdot ( \rho \vec v ) = 0 \\
\displaystyle \partial_t ( \rho \vec v ) + \nabla \cdot ( \rho \vec v \vec v + ( P + \frac12 B^2 ) \tensor{I} - \vec B \vec B ) = 0 \\
\displaystyle \partial_t ( \frac12 \rho v^2 + \epsilon + \frac12 B^2 ) + \\
\displaystyle \nabla \cdot ( ( \frac12 \rho v^2 + \epsilon + P + B^2 ) \vec v - (v \cdot B) \vec B ) = 0 \\
\displaystyle \partial_t ( \vec B ) + \nabla \cdot ( \vec v \vec B - \vec B \vec v ) = 0
\end{array} 
\label{eqn:bare}
\end{equation}

Rewriting the equations in a cylindrical basis, it is possible to express these as additions to the standard hydro variables:

\begin{equation}
u = u_{\rm hydro} + u_{\rm mhd},
\end{equation}
where the $u_{\rm hydro}$ are given by (\ref{eqn:bareu}), and

\begin{equation}
u_{\rm mhd} = \{ 0 , 0 , 0 , 0 , \frac12 B^2 , B_r , B_\phi/r , B_z \}.
\label{eqn:mhdcons}
\end{equation}

The MHD fluxes are similarly summarized as

\begin{eqnarray}
\vec F_{\rm mhd} = \{ 0 , r( \frac12 B^2 \hat \phi - B_\phi \vec B ) , \frac12 B^2 \hat r - B_r \vec B ,  \nonumber \\
\frac12 B^2 \hat z - B_z \vec B , B^2 \vec v - v \cdot B \vec B , \nonumber \\
B_r \vec v - v_r \vec B , (B_\phi \vec v - v_\phi \vec B)/r , B_z \vec v - v_z \vec B\}.
\label{eqn:mhdflux}
\end{eqnarray}
and there is a new source term for radial momentum, given by magnetic tension, or ``hoop stress":

\begin{equation}
S_{\rm mhd} = \{ 0 , 0 , B_{\phi}^2/r , 0 , 0 , 0 , 0 , 0 \}.
\label{eqn:mhdsrc}
\end{equation}

MHD is a challenging set of field equations to integrate, because of subtle behaviors relating to the divergence constraint, $\nabla \cdot B = 0$ \citep[e.g.][]{1980JCoPh..35..426B, 1988ApJ...332..659E, 1998ApJ...494..317D, 1998ApJ...509..244R, 1999JCoPh.149..270B, 2000JCoPh.161..605T}.

One way of framing the ${\nabla \cdot B}$ problem is in terms of the following thought experiment: consider a two-dimensional cartesian domain with a field loop advecting with a uniform velocity in the x direction.  While $B_y$ is advected, there is no x-directed flux for the parallel component $B_x$.  Instead, the advection of $B_x$ is accounted for by the rotational flux in the $y$ direction.
\begin{equation}
\partial_t B_x - \partial_y( v_x B_y ) = 0
\end{equation}
but the divergence constraint implies
\begin{equation}
\partial_y B_y = - \partial_x B_x
\end{equation}
so if ${\nabla \cdot B}$ is guaranteed to be zero by the numerical stencil which updates the magnetic field, then the rotational flux in y is identical to an advective flux in the x direction, so it makes no difference that ${B_x}$ is not explicitly advected in this sense.

However, if one cannot guarantee that ${\nabla \cdot B = 0}$ then something as simple as advecting a field loop can go wrong.  ${B_x}$ and ${B_y}$ are numerically updated in a fundamentally different way, and this causes the loop to eventually destabilize.  This becomes more of a problem the larger ${\nabla \cdot B}$ is allowed to grow.




Many techniques have been developed to stably evolve this system, the most successful of which is constrained transport \citep[CT,][]{1988ApJ...332..659E}.  However, CT is often difficult to implement on complicated grids, in particular moving meshes such as DISCO's (though notably, \cite{2014MNRAS.442...43M} employed CT on a 2D Voronoi mesh, demonstrating that CT is possible even with complex geometries).  As a result, several methods have been developed which have less dependence on the mesh employed.  One popular such technique is to modify the evolution equations so that the divergence constraint propagates and diffuses \citep[e.g.][]{2002JCoPh.175..645D}.

Such divergence-cleaning techniques are easy to implement, but it is difficult to test their effectiveness, because they do not guarantee machine-precision zero divergence for any stencil.  Even a small divergence error can cause inaccurate physics on long timescales.

Another issue with the formulation of \cite{2002JCoPh.175..645D} is that it introduces an additional wavespeed into the system.  In the standard formulation, this wavespeed must be the fastest velocity in the system, in order that these waves can keep up with divergence errors quickly enough to correct them.  Unfortunately for the moving mesh technique, this eliminates the time-step advantage, because this wave moves quickly with respect to the grid.

Alternate Galilean-invariant formulations of these equations are possible \citep[e.g.][]{1999JCoPh.154..284P}, but these always require source terms which have derivatives, which undermines many of the advantages of the finite-volume formulation (one principal advantage of finite volume methods is that they evolve the integral form of the equations, and as a result do not necessitate smooth solutions).  In short, it may be more advantageous to prevent divergence errors from appearing in the first place.

Another method increasingly employed is the vector potential formulation \citep{2003AnA...400..397D, 2010PhRvD..82h4031E}.  This, of course, has the advantage that divergence errors are never introduced, as B is defined as a curl.  It is also a preferred method for nontrivial meshes, which can complicate implementations of CT.  Some vector potential formulations have been shown to be functionally equivalent to CT on uniform meshes \citep{2010PhRvD..82h4031E, 2011JCoPh.230.3803H, helzel2013high}.

Unfortunately, the vector potential formulation might also require a numerical derivative, in the operation $B = \nabla \times A$, and therefore some formulations might assume that the vector potential is differentiable (for example, this is generally the case if the vector potential is cell-centered).  Another disadvantage is that introducing a vector potential also introduces gauge modes.  Either these gauge modes are static, and accumulate on the grid, or they propagate at some velocity which must be introduced, and they can also limit the code's time-step, similar to Dedner's method.  The time-step advantage is big enough for the moving mesh that it is worth maintaining, if possible.  Of course, none of these disadvantages can be truly devastating for vector potential formulations, as any constrained transport scheme could be re-expressed as a vector potential scheme, by evolving $\vec A$ on each edge of the mesh using the same electric fields, and calculating the magnetic flux through each face by integrating $A$ around a closed loop.  In this case, the truncation error can be identical to a CT scheme.

Constrained transport techniques find ways of processing the MHD fluxes so that they do not introduce any divergence errors.  In a sense, the idea is somewhat analogous to conservative formulations, which evolve the system in such a way as to avoid conservation-law violations.

The CT formulation of \cite{1988ApJ...332..659E}, the most commonly employed CT scheme, makes use of the natural topology of the MHD equations.  Faraday's law can be expressed as a conservation law, and it is straightforward to define conserved quantities in volumes, and fluxes through faces.  On the other hand, $\vec B$ is a \textit{conserved flux}.  In other words, Faraday's law is more naturally expressed in a lower-dimensional form, by integrating it over the area of a face:

\begin{equation}
\partial_t \Phi = - \oint \vec E \cdot \vec{dl}.
\label{eqn:faraday}
\end{equation}

This is a lower-dimensional analog to the finite-volume conservation law (equation \ref{eqn:intcons}).  The magnetic flux $\Phi$ is the analog of mass, and the electric field $E$ is the analog of mass flux (the direction of the flow is perpendicular to the electric field, but the component of $E$ along the edge provides the magnitude of the flux).

Given a computed value of $\vec E = - \vec v \times \vec B $, one can generalize the integral equations derived in section \ref{sec:integral}, to arrive at the ``finite-area" form of (\ref{eqn:faraday}):

\begin{equation}
\Phi^{n+1}_i = \Phi^n_i - \Delta t \sum\limits_{\rm edge ~e} \vec {dl}_{e} \cdot \vec E_{e}.
\label{eqn:far1}
\end{equation}

\begin{figure}
\epsscale{1.0}
\plotone{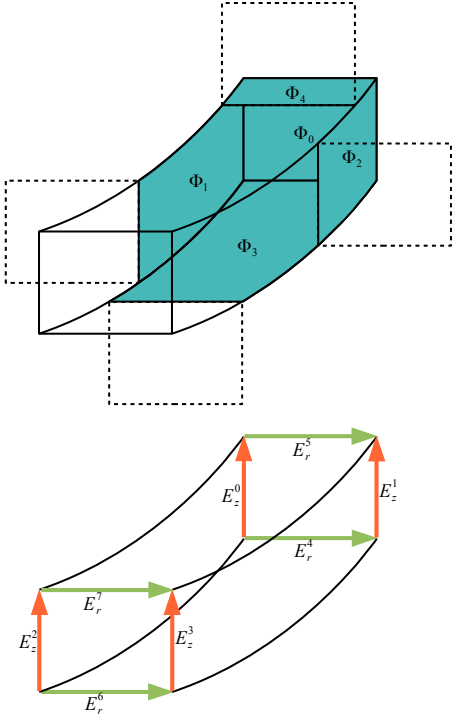}
\caption{ Upper Panel: Zone-specific faces are identified.  Each zone can be uniquely associated with five faces in three dimensions (or three faces in two dimensions).  Lower Panel: Zone-specific edges are identified.  Radial and vertical edges are associated with the interior a given zone.  Some of these edges are redundant, as they are duplicated in other zones.  This is accounted for by averaging the electric field between associated edges on neighboring zones.
\label{fig:facedge1} }
\end{figure}

The numerical scheme can be interpreted as a lower-dimensional analog of standard finite-volume methods, with $\Phi$ playing the role of mass, and $E$ being the analog of mass flux (see Figure \ref{fig:eflux1}).  Because the fluxes $\Phi$ are stored on faces, this requires some memory of the numerical mesh to persist from one time-step to the next.  For convenience, each flux $\Phi$ is stored on one of the zones which houses the face associated with this flux (see Figure \ref{fig:facedge1}).  The 3D CT scheme used in DISCO is based on the method of \cite{1988ApJ...332..659E}, though the electric fields are not computed in exactly the same way.

\subsubsection{Face-Centered Fluxes to Cell-Centered Fields}

The complete description of the time-evolution for the magnetic fluxes will appear in section \ref{sec:mhdtime}.  First, it is important to recognize that it will not be possible to calculate the magnetic field directly from the magnetic flux.  This would require the operation $\left< B \right> = \Phi / dA$, where $dA$ is the area of the face.  Unfortunately, this area can be arbitrarily small, as faces can disappear or reappear during changes in the topology of the mesh.  Calculating the magnetic field by dividing a small flux by a small area can result in arbitrarily large errors; small errors in the flux would lead to arbitrarily large errors in the field.  Therefore, in DISCO's numerical scheme the operation $\left< B \right> = \Phi / dA$ is never explicitly performed.  Instead, the fluxes on each face are used to determine a cell-centered average magnetic field, and this magnetic field is used to update the system.  In fact, most of the timestep proceeds as a normal finite-volume scheme, treating the cell-centered magnetic fields as primitive variables; they are interpolated to the faces along with the other variables, and used as input to Riemann solvers.  The only time the face-centered fluxes are needed is in the step just before converting from conserved to primitive variables, where the cell-centered magnetic fields are determined based on the face-centered fluxes.  In this sense, one could interpret the update of the magnetic fields as a predictor-corrector scheme \citep{2011JCoPh.230.3803H}.

This operation (from $\Phi_{\rm face}$ to $B_{\rm zone}$) is performed by determining the total flux through each surface of the zone; the flux is then assumed to vary linearly through the zone, so that

\begin{equation}
\langle \vec B \cdot \hat n \rangle_{\rm cell} dA_c = \frac12 ( \sum_{\rm side~1} \Phi + \sum_{\rm side~2} \Phi ).
\end{equation}

In other words, the cell-centered components of the magnetic field are

\begin{equation}
\langle B_r \rangle_{\rm cell} = { ( \sum \Phi_{\rm in} + \sum \Phi_{\rm out} ) \over ( r_+ + r_- ) \Delta \phi \Delta z }
\end{equation}

\begin{equation}
\langle B_z \rangle_{\rm cell} = { ( \sum \Phi_{\rm top} + \sum \Phi_{\rm bottom} ) \over ( r_+ + r_- ) \Delta \phi \Delta r }
\end{equation}

\begin{equation}
\langle B_\phi \rangle_{\rm cell} = { ( \Phi_{\rm front} + \Phi_{\rm back} ) \over 2 \Delta r \Delta z }
\end{equation}

Once these zone-centered fields are calculated, the conserved variables in the zone are converted to primitive variables, using the zone-centered magnetic fields just computed.  The rest of the time-update proceeds using these fields as cell-centered primitive variables.

\subsubsection{Time-Update of the Magnetic Fluxes}
\label{sec:mhdtime}

In section \ref{sec:getefields}, it will be explained how the electric field is calculated on each edge.  For now, it is assumed that this $E$ field is known, and the goal is to derive a rule to update the magnetic fluxes during a timestep.  This starts with Faraday's law:

\begin{equation}
\partial_t \vec B + \nabla \times \vec E = 0
\end{equation}

Following the same process which led to (\ref{eqn:evolution}), Faraday's law is integrated over the surface of a face:

\begin{equation}
\partial_t \Phi + \oint \vec E \cdot dl = 0
\end{equation}

Now, interpreting $\vec E_e$ as the time-averaged electric field on edge $e$, the time-update step can be expressed as:

\begin{equation}
\Phi^{n+1}_f = \Phi^n_f - \Delta t \sum\limits_{\rm edge ~e} \vec E_{e} \cdot dl_{e}.
\end{equation}

This method of evolving face-centered fluxes does not conserve volume-integrated flux density, as some uniform-mesh CT techniques do \citep[e.g.][]{2000JCoPh.161..605T}.  If the edges move with velocity $\vec w$, the generalization is:

\begin{equation}
\Phi^{n+1}_f = \Phi^n_f - \Delta t \sum\limits_{\rm edge ~e} ( \vec E_{e} + \vec w_{e} \times \vec B_{e} ) \cdot dl_{e}.
\end{equation}

This is also how mesh motion is accounted for by \cite{2014MNRAS.442...43M}.  Just like equation (\ref{eqn:evolution}), this is an exact expression, suitably interpreted.  All of the numerical approximations will be housed in the calculation of $E_{e}$ and $B_{e}$, the time-averaged electric and magnetic fields.  These will be determined in the following section.

\subsubsection{Calculation of the Electric Fields}
\label{sec:getefields}



Any choice for the electric fields will preserve the divergence constraint, but it is important to make a well-motivated choice in order to provide an accurate and stable approximation to the underlying field equations.  For example, \cite{1999JCoPh.149..270B} showed that simply averaging the electric fields on four adjacent faces fails to produce an upwind scheme, and as a result this can give innacurate solutions.  Various remedies are proposed for this \citep[e.g.][]{1999JCoPh.149..270B, 2005JCoPh.205..509G}.  Here, the numerical scheme is not as susceptible to this upwinding issue, because the mesh motion typically keeps azimuthal faces within the Riemann fan.  However, it is still nontrivial to develop a stable scheme because of the complicated mesh structure.  In practice, a stable scheme was found by experimenting with different averaging procedures.  The calculation of the electric fields will involve three steps: first, a definition of zone-specific edge-centered fields, secondly an identification of these electric fields with fluxes found in the Riemann solver step, and finally an averaging process over several neighboring zones.

Indexing of the faces and edges in each zone is shown in Figure \ref{fig:facedge1}.  To calculate the electric field on these edges, note that this electric field gives the number of magnetic field lines being dragged across each edge per unit time.  If one knows, for example, the flow of $B_\phi$ through a face with r-directed normal, then one implicitly knows the line integral of $E_z$ along a z-directed edge parallel to that face.

These statements can be made more mathematically concrete.  Using the geometrical fact that the area element is given by the wedge product of two line elements $dA_i = \frac12 \epsilon_{ijk} dx_j \wedge dx_k$, the integrated flux of a magnetic field component $B_i$ through a face is given by

\begin{equation}
F(B_i) \cdot dA = (v_j B_i - v_i B_j) \frac12 \epsilon_{jkl} dx_k \wedge dx_l
\end{equation}

The combination $(v_j B_i - v_i B_j)$ can be related to the cross product $\vec v \times \vec B$:

\begin{equation}
F(B_i) \cdot dA = \epsilon_{jim} (v \times B)_m \frac12 \epsilon_{jkl} dx_k \wedge dx_l
\end{equation}

Given $\vec E = - \vec v \times \vec B$ and the identities of the $\epsilon$ pseudotensor, one can arrive at the relation:

\begin{equation}
F(B_i) \cdot dA = E_j dx_j \wedge dx_i.
\end{equation}

If one integrates this formula over a face, then divides by area, one attains

\begin{equation}
\left< F(B_i) \right> = { \int E \cdot dl \wedge dx_i \over dA},
\label{eqn:eflux}
\end{equation}
where brackets denote area-averaged fluxes.

Therefore, the line-integrated electric field can be related to the area-integrated flux of magnetic field, which has already been calculated from the Riemann solver.  In other words, one can associate a line-averaged $E$ with an area-averaged flux.  For each electric field component, there are two areas to average over, corresponding to the two coordinate planes parallel to this edge.  The electric field is set to the mean of these two area-averaged fluxes.  These fields are not averaged with neighboring zones yet, as this will be done in the next step.  Specifically, for the vertical fields,

\begin{eqnarray}
E_{z}^0 &=& \frac12 \left< \vec F_{B_r} \cdot \hat \phi \right>_{\rm front~face} - \frac12 \left< \vec F_{B_\phi} \cdot \hat r \right>_{\rm inner~face} \\
\label{eqn:beginE}
E_{z}^1 &=& \frac12 \left< \vec F_{B_r} \cdot \hat \phi \right>_{\rm front~face} - \frac12 \left< \vec F_{B_\phi} \cdot \hat r \right>_{\rm outer~face} \\
E_{z}^2 &=& \frac12 \left< \vec F_{B_r} \cdot \hat \phi \right>_{\rm back~face} - \frac12 \left< \vec F_{B_\phi} \cdot \hat r \right>_{\rm inner~face} \\
E_{z}^3 &=& \frac12 \left< \vec F_{B_r} \cdot \hat \phi \right>_{\rm back~face} - \frac12 \left< \vec F_{B_\phi} \cdot \hat r \right>_{\rm outer~face} 
\end{eqnarray}
where ``front" implies the face with greater value of $\phi$.  For the radial fields,

\begin{eqnarray}
E_{r}^4 &=& \frac12 \left< \vec F_{B_\phi} \cdot \hat z \right>_{\rm bottom~face} - \frac12 \left< \vec F_{B_z} \cdot \hat \phi \right>_{\rm front~face} \\
E_{r}^5 &=& \frac12 \left< \vec F_{B_\phi} \cdot \hat z \right>_{\rm top~face} - \frac12 \left< \vec F_{B_z} \cdot \hat \phi \right>_{\rm front~face} \\
E_{r}^6 &=& \frac12 \left< \vec F_{B_\phi} \cdot \hat z \right>_{\rm bottom~face} - \frac12 \left< \vec F_{B_z} \cdot \hat \phi \right>_{\rm back~face} \\
E_{r}^7 &=& \frac12 \left< \vec F_{B_\phi} \cdot \hat z \right>_{\rm top~face} - \frac12 \left< \vec F_{B_z} \cdot \hat \phi \right>_{\rm back~face}.
\label{eqn:endE}
\end{eqnarray}

Azimuthal electric fields will be treated separately, but will have the same essential forms:

\begin{equation}
E_{\phi} = \frac12 \left< \vec F_{B_z} \cdot \hat r \right> - \frac12 \left< \vec F_{B_r} \cdot \hat z \right>
\end{equation}

This choice of electric field is designed for consistency of the numerical scheme, since the flux of field lines is consistent with the Godunov fluxes calculated from the Riemann solver.

After calculating these zone-specific edge-centered electric fields, an averaging process is performed between zones in order to acquire a well-posed electric field on each individual edge in the mesh.

\subsubsection{Self-Consistent Averaged Electric Fields}
\label{sec:avg}

Once zone-specific electric fields have been specified via equations (\ref{eqn:beginE}) - (\ref{eqn:endE}), an averaging process is performed, in order to ensure that electric fields on adjacent faces are consistent with one another.  For example, as Figure \ref{fig:avg1} shows, $E_z^0$ on zone $i$ should be identical to $E_z^2$ on zone $i+1$.  Before the averaging process, these fields are not generally consistent with one another.  After averaging, the electric field is self-consistent and the MHD update stencil effectively spans a larger number of zones.

As Figure \ref{fig:avg1} demonstrates, $E^0_z$ and $E^1_z$ of zone $i$ must be compatible with $E^2_z$ and $E^3_z$ of zone $i+1$.  This is accomplished via the substitution:

\begin{figure}
\epsscale{1.0}
\plotone{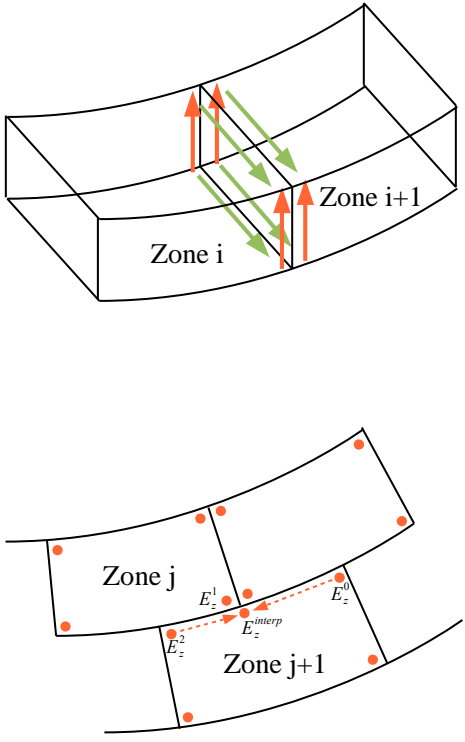}
\caption{ Neighboring zones have inconsistent electric fields.  These electric fields are made consistent by replacing each with the average of the two.  In the upper panel, edges are identified between zones which share a $\phi$-normal face (equations \ref{eqn:figavg1}-\ref{eqn:figavg2} and \ref{eqn:figavg3}-\ref{eqn:figavg4}).  In the lower panel, the interpolation procedure is shown (corresponding to equation \ref{eqn:figavg}), to identify edges in zones on neighboring annuli.
\label{fig:avg1} }
\end{figure}

\begin{figure}
\epsscale{1.0}
\plotone{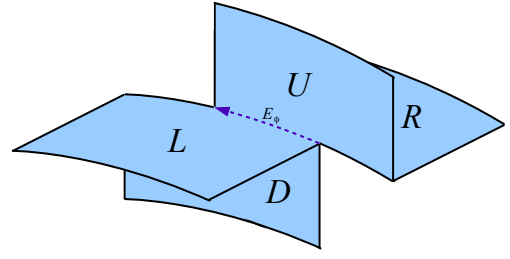}
\caption{ Azimuthal electric fields are found by averaging over four adjacent faces.  The resultant electric field is calculated in equation (\ref{eqn:e__phi}).
\label{fig:avg2} }
\end{figure}

\begin{eqnarray}
E_{z~i}^0 \rightarrow \frac12 ( E_{z~i}^0 + E_{z~i+1}^2 ) \label{eqn:figavg1}\\
E_{z~i}^1 \rightarrow \frac12 ( E_{z~i}^1 + E_{z~i+1}^3 ) \\
E_{z~i}^2 \rightarrow \frac12 ( E_{z~i}^2 + E_{z~i-1}^0 ) \\
E_{z~i}^3 \rightarrow \frac12 ( E_{z~i}^3 + E_{z~i-1}^1 ) \label{eqn:figavg2}
\end{eqnarray}

Similarly, $E^0_z$ of zone $j$ must be compatible with the interpolated electric field of zone $j+1$:

\begin{eqnarray}
E_{z~j}^0 \rightarrow \frac12 ( E_{z~j}^0 + E_{z~j-1}^{\rm interp} ) \\
E_{z~j}^1 \rightarrow \frac12 ( E_{z~j}^1 + E_{z~j+1}^{\rm interp} ) \label{eqn:figavg}\\
E_{z~j}^2 \rightarrow \frac12 ( E_{z~j}^2 + E_{z~j-1}^{\rm interp} ) \\
E_{z~j}^3 \rightarrow \frac12 ( E_{z~j}^3 + E_{z~j+1}^{\rm interp} )
\end{eqnarray}
where $E_{z_j+1}^{\rm interp}$ for example is a weighted average of $E^0_z$ and $E^2_z$ interpolated to the position of $E^1_{z~j}$ or $E^3_{z_j}$ (Figure \ref{fig:avg1}).  Similarly, for the radial electric fields:

\begin{eqnarray}
E_{r~i}^4 \rightarrow \frac12 ( E_{r~i}^4 + E_{r~i+1}^6 ) \label{eqn:figavg3}\\
E_{r~i}^5 \rightarrow \frac12 ( E_{r~i}^5 + E_{r~i+1}^7 ) \\
E_{r~i}^6 \rightarrow \frac12 ( E_{r~i}^6 + E_{r~i-1}^4 ) \\
E_{r~i}^7 \rightarrow \frac12 ( E_{r~i}^7 + E_{r~i-1}^5 ) \label{eqn:figavg4}\\
E_{r~k}^4 \rightarrow \frac12 ( E_{r~k}^4 + E_{r~k-1}^{\rm interp} ) \\
E_{r~k}^5 \rightarrow \frac12 ( E_{r~k}^5 + E_{r~k+1}^{\rm interp} ) \\
E_{r~k}^6 \rightarrow \frac12 ( E_{r~k}^6 + E_{r~k-1}^{\rm interp} ) \\
E_{r~k}^7 \rightarrow \frac12 ( E_{r~k}^7 + E_{r~k+1}^{\rm interp} )
\end{eqnarray}

$E_\phi$ is not averaged with neighbors in this way.  Rather, the $\phi$ component of the electric field is not defined zone-wise, but on the intersection between four faces (Figure \ref{fig:avg2}).  In this case,

\begin{eqnarray}
E_{\phi} = \frac14 ( \left< F(B_z) \cdot \hat r \right>_{U} + \left< F(B_z) \cdot \hat r \right>_{D} \\
- \left< F(B_r) \cdot \hat z \right>_{L} - \left< F(B_r) \cdot \hat z \right>_{R} )
\nonumber
\label{eqn:e__phi}
\end{eqnarray}

Note that on average each zone is in contact with $16$ azimuthal edge segments, and therefore $E_{\phi}$ is effectively averaged over more than four faces.

\subsubsection{Topology Changes}

Because the zones shear past one another, faces can disappear or emerge during the course of a timestep (Figure \ref{fig:flip}).  Ordinarily, this might be a problem for the numerical method; when a face disappears, where does its flux go?  When a face emerges, what sets its flux?

The way this scheme is designed, the numerical solution is not very sensitive to the answers to these questions, since the area of the face in question is very small in these circumstances, and the cell-centered magnetic field is an average over faces, weighted by face area.  Nonetheless, some flux must be specified.  Since the face is changing topology, its bounding edges must be very close together and therefore the electric and magnetic fields must be very similar (this is guaranteed by the interpolation and averaging process).  Therefore, most of the time-dependence of the flux must come from the difference in the velocity $\vec w$ at each edge:

\begin{equation}
\Delta \Phi \approx ( w_L - w_R ) B dl dt.
\label{eqn:shortphi}
\end{equation}

In other words, during a topology flip the change in flux in the cell is assumed to be entirely due to the amount of flux overtaken by the edge.  In this case, during the time update, there needs to be an adjustment in order to account for the fact that in the first part of the timestep, the face was giving up its flux to one zone, and in the last part of the timestep, it was taking flux from another zone.

Note that because $|w_L - w_R| dl dt > dA$ during a time-step where the topology flips, Equation (\ref{eqn:shortphi}) guarantees a sign change in $\Phi$ during a topology change; the point at which $\Phi$ goes to zero signifies the topology flip.  Therefore, in Figure \ref{fig:flip} all flux lost before the sign change should be given to face 1 and all flux accumulated after this point should be taken from face 2.

Algorithmically, all of this is simple to account for.  First, a normal timestep is taken.  Then, after the timestep, if a face ``flips", i.e. a face disappears from one zone and moves onto another zone, then the following corrections must be made to this face and the adjacent faces (see Figure \ref{fig:flip}):

\begin{figure}
\epsscale{1.0}
\plotone{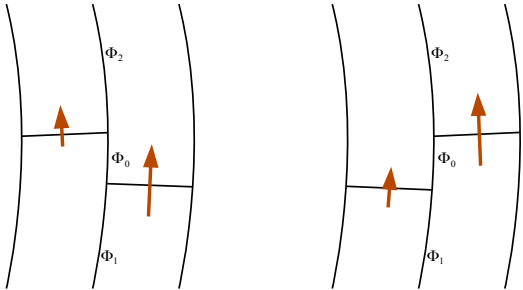}
\caption{ Schematic diagram of a topology flip in the mesh.  After a topology flip, magnetic fluxes $\Phi_0$, $\Phi_1$ and $\Phi_2$ are adjusted for the new topology using equations (\ref{eqn:top1} - \ref{eqn:top3})
\label{fig:flip} }
\end{figure}

\begin{eqnarray}
\Phi^{\rm new}_{0} &=& -\Phi_{0} \label{eqn:top1}\\
\Phi^{\rm new}_1 &=& \Phi_1 + \Phi_{0} \\
\Phi^{\rm new}_2 &=& \Phi_2 + \Phi_{0} \label{eqn:top3}
\end{eqnarray}

This adjustment also guarantees that the magnetic divergence remains zero for the appropriate stencil.  This method for tracking topology changes is different from \cite{2014MNRAS.442...43M}, where flux is redistributed equally between neighboring faces. 

\section{Test Problems}
\label{sec:tests}

Given the uniqueness of this numerical scheme, a wide range of tests is performed.  Many of these tests are ``sanity checks", ensuring that all terms are correctly accounted for in the code.  Accuracy and convergence are also very important, and therefore several convergence tests are performed.  Additionally, several challenging tests relevant to astrophysics are performed, in order to demonstrate the code's robustness and usefulness for studying nontrivial astrophysical flows, including tests of disk-planet interactions, and MHD turbulence driven by the magnetorotational instability in an accretion flow.

Nearly all tests use an adiabatic index of $\gamma = 5/3$, or an isothermal equation of state.  The two exceptions are the MHD rotor test, which uses an adiabatic index of $\gamma = 1.4$, and the supersonic Keplerian spreading test, which uses a nearly isothermal adiabatic index $\gamma = 1.001$.  Unless otherwise specified, all hydro calculations use the HLLC Riemann solver, and all MHD calculations use the HLLD Riemann solver.  The slope-limiting parameter $\theta_{\rm plm} = 1.5$ in all tests, and the CFL number is $0.5$ for all hydro tests, and $0.2$ for all MHD tests.

\subsection{Hydrodynamics}

\subsubsection{Cylindrical Shock Tube}

\begin{figure}
\epsscale{1.0}
\plotone{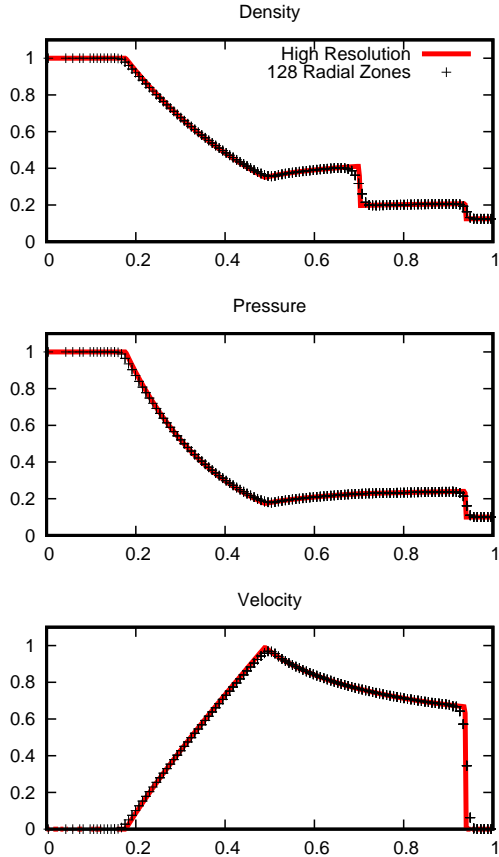}
\caption{ Cylindrical shock tube test at $t = 0.25$.  DISCO's grid with 128 radial zones is compared with an identical high-resolution calculation using a 1D code in cylindrical coordinates.  Qualitative agreement is found; DISCO accurately captures the shocks propagating radially, even though mesh motion is not utilized in this test.
\label{fig:shocktube} }
\end{figure}

This test demonstrates DISCO's ability to capture radially propagating shocks. For characteristics moving radially, DISCO performs as a robust high-resolution shock-capturing code, even when the mesh motion is not utilized.

The domain extends from $0 < r < 1$, with uniform radial resolution.  An adiabatic index of $\gamma=5/3$ is employed.  If $r<0.5$, $\rho=1.0$ and $P=1.0$.  Otherwise, $\rho=0.125$ and $P=0.1$.  Initially the fluid is not moving: $v_r = \omega = 0$.  This calculation is run until a time $t=0.25$.

In Figure \ref{fig:shocktube} density, pressure, and velocity are plotted at this time, using a resolution of 128 radial zones.  This is compared with a high-resolution calculation using a 1D code in cylindrical coordinates.  Qualitative agreement is illustrated in Figure \ref{fig:shocktube}.

\subsubsection{Cylindrical Isentropic Pulse}

\begin{figure}
\epsscale{1.0}
\plotone{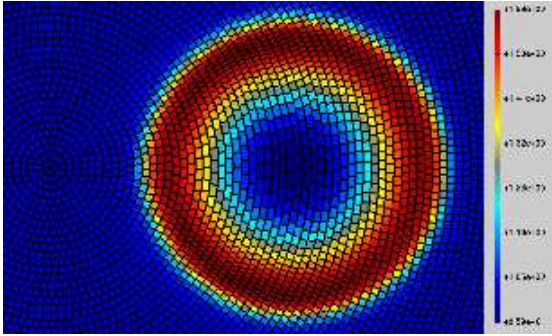}
\caption{ Cylindrical isentropic pulse test at $t = 0.1$ using $N_r = 64$ radial zones.  The pulse is offset with respect to the origin to test the convergence of a problem which includes all hydro fluxes.  In this test, each zone is moved individually, so that zones are compressed or expanded with the flow.
\label{fig:isentropic1} }
\end{figure}

\begin{figure}
\epsscale{1.0}
\plotone{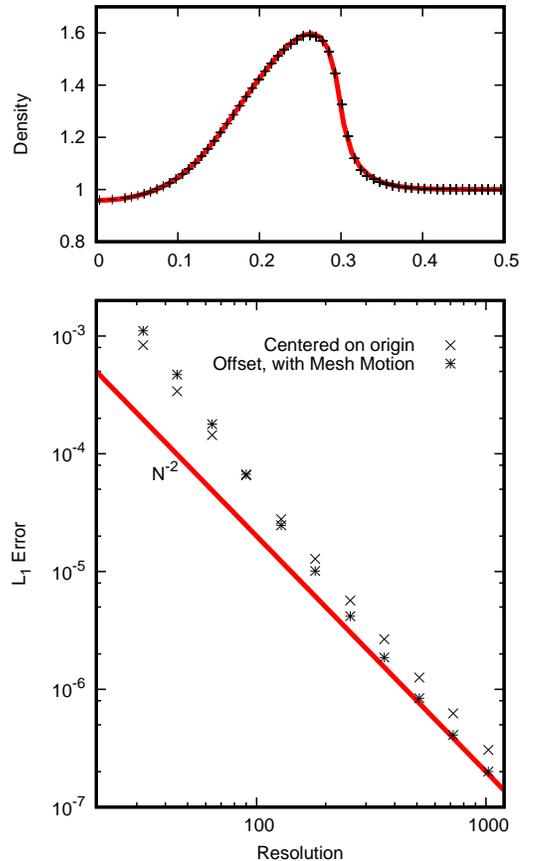}
\caption{ Convergence of the cylindrical isentropic pulse test.  The upper panel compares the run with $N_r = 64$ with a high-resolution calculation with a 1D code in cylindrical coordinates.  The lower panel measures the $L_1$ error of this solution, testing entropy conservation (\ref{eqn:l1_entropy}), showing DISCO's second-order convergence on this problem, including when the pulse is offset with respect to the origin and mesh motion is utilized.
\label{fig:isentropic2} }
\end{figure}

\begin{figure*}
\epsscale{1.0}
\plotone{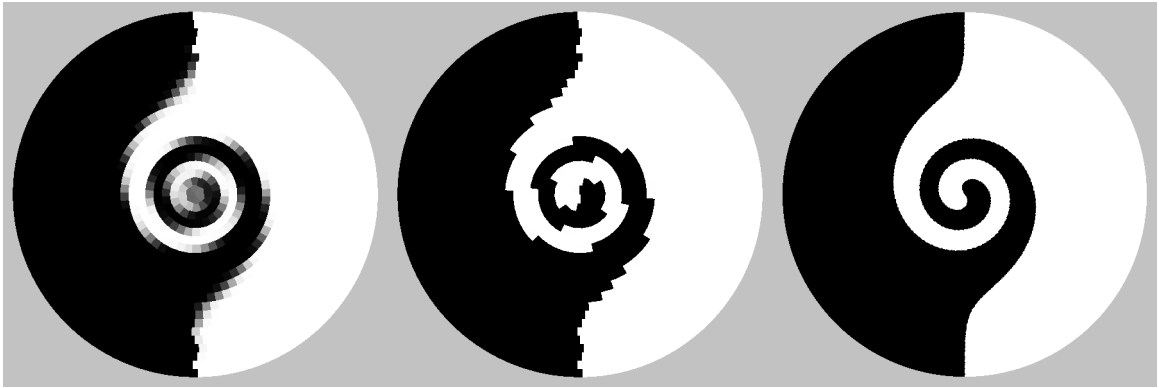}
\caption{ Passive scalar in the smooth vortex test at $t = 10$, demonstrating DISCO's ability to maintain contact discontinuities to high precision.  The first two panels use $N_r = 64$, with fixed mesh (left) and moving mesh (center).  The fixed mesh diffuses out the passive scalar, whereas the moving mesh preserves the contact discontinuity precisely.  A higher-resolution run ($N_r = 256$) is shown in the final panel for comparison.
\label{fig:vortex1} }
\end{figure*}

\begin{figure}
\epsscale{1.0}
\plotone{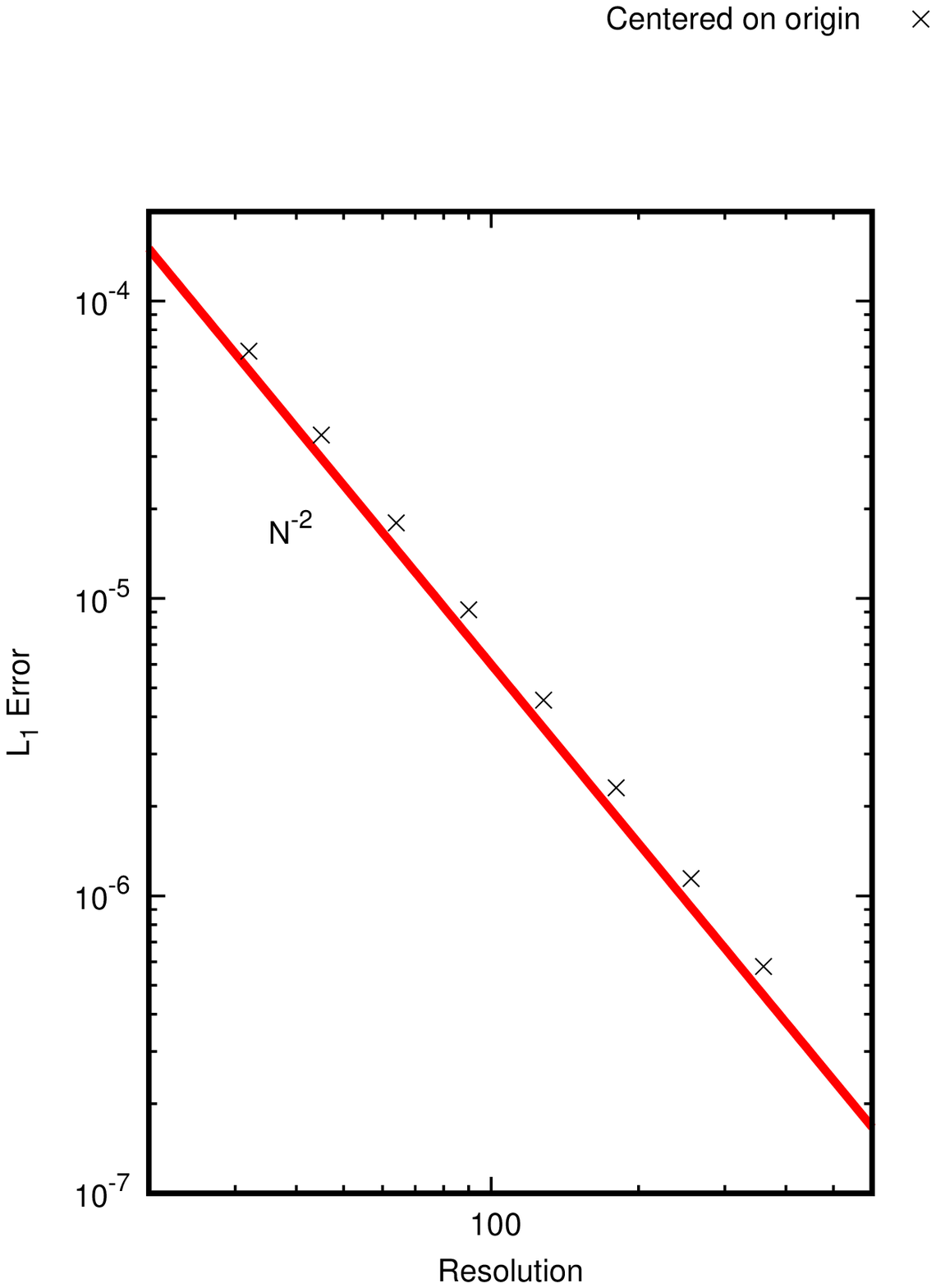}
\caption{ Convergence of the smooth vortex test.  Error is given by (\ref{eqn:l1_density}) and resolution indicates the number of radial zones.  DISCO converges at second-order for this test.
\label{fig:vortex2} }
\end{figure}

The isentropic pulse is a simple nonlinear test for convergence of any code.  Convergence is checked by measuring entropy conservation. The initial setup is as follows:

The domain extends from $0<r<0.5$, with uniform radial resolution. Density is given by

\begin{equation}
\rho = 1 + 3 e^{-(80 r^2)}.
\end{equation}

Pressure is chosen to be isentropic:

\begin{equation}
P = \rho^\gamma
\end{equation}
and the fluid is initially stationary, $v_r = \omega = 0$. The pulse explodes outward (Figure \ref{fig:isentropic1}), and eventually forms a shock, but before the shock forms the equation $P = K \rho^\gamma$ continues to hold due to entropy conservation ($K = P/\rho^{\gamma}$ evolves as a conserved scalar, as long as the solution remains smooth).

Error is calculated by verifying entropy conservation at a time $t=0.1$, before the shock has formed:

\begin{equation}
L_1 = { \int |P/\rho^{5/3}-1.0|dV \over \int dV}.
\label{eqn:l1_entropy}
\end{equation}

Fast convergence is found for this problem (Figure \ref{fig:isentropic2}); for resolutions lower than $1024$, convergence is faster than second order. At higher resolutions, convergence is second order.

This test is effectively 1D, as all motion is radial; it can be run with arbitrarily low angular resolution.  However, it can also be used as a multidimensional test problem, by offsetting the origin of the pulse.  Additional calculations with these initial conditions were run with an origin offset by $\Delta y=0.5$ (this is the example shown in Figure \ref{fig:isentropic1}).  Convergence is also second order in this offset example, when each zone is allowed to move independently (Figure \ref{fig:isentropic2}).

\subsubsection{Smooth Vortex}

This test helps to demonstrate convergence, and the effectiveness of the DISCO code at preservation of contact discontinuities.  The grid has uniform radial resolution from $0<r<5$.  Density is uniform with $\rho = 1.0$, and angular velocity is chosen as

\begin{equation}
\omega(r)=e^{-\frac12 r^2}
\end{equation}

Pressure is chosen to balance centrifugal forces: $\rho \Omega^2 r= \partial_r P$.  This results in the following pressure:

\begin{equation}
P(r)=1-\frac12 e^{-r^2}.
\end{equation}

The vortex is trans-sonic (maximum Mach number of about 0.53). In Figure \ref{fig:vortex1}, a passive scalar is included to demonstrate the code's ability to maintain contact discontinuities and prevent artificial diffusion.  When mesh motion is turned off, the contact discontinuity is smeared out.  With mesh motion turned on, the contact discontinuity is maintained precisely.  The first two panels use a low resolution of 64 zones.  A calculation at a resolution of 256 radial zones is also included for comparison.

Error is calculated at $t=10$ using the density:

\begin{equation}
L_1 = { \int |\rho-1.0|dV \over \int dV}.
\label{eqn:l1_density}
\end{equation}

Figure \ref{fig:vortex2} shows clear second-order convergence on this test.

\subsubsection{Supersonic Keplerian Shear Flow}
\label{sec:kepler}

\begin{figure}
\epsscale{1.0}
\plotone{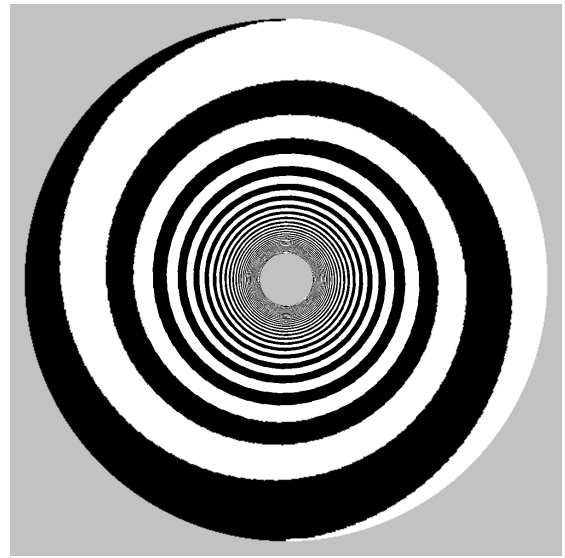}
\caption{ Passive scalar in the Keplerian shear flow test, demonstrating DISCO's ability to precisely preserve Keplerian orbital flow, and to preserve contact discontinuities to high precision.  The passive scalar is plotted after a single orbit at the outer boundary, corresponding to roughly $32$ orbits at the inner boundary.  Preserving a Keplerian shear flow accurately is essential for any code which is being used to study disks, in order that whatever phenomenon being studied is not swamped out by errors from this background flow.
\label{fig:kepler1} }
\end{figure}

\begin{figure}
\epsscale{1.0}
\plotone{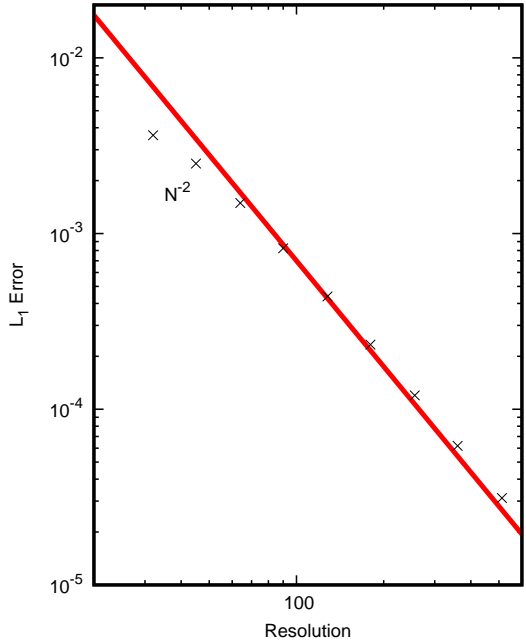}
\caption{ Convergence of the Keplerian shear flow test.  Error is given by (\ref{eqn:l1_density}) and resolution indicates the number of radial zones.  DISCO achieves second-order convergence.
\label{fig:kepler2} }
\end{figure}

This is an important test, as most problems DISCO was designed to solve have a Keplerian background flow. This stationary flow must be captured accurately if one wishes to study some subtle phenomenon which is a perturbation to this flow.

The setup for this problem is as follows.  Zones are logarithmically spaced in the range $0.1<r<1.0$.  The density and pressure are uniform with $\rho = 1.0$, $P = 0.01$.  Velocity is Keplerian: $v_r = 0$, $\omega(r) = r^{-3/2}$.

Boundary conditions are fixed at these initial conditions. A point mass is inserted at $r = 0$ so that centrifugal and gravitational forces are balanced.  For this problem, the gravitational potential was not smoothed, since the point mass is not on the grid, and the Keplerian flow is an exact solution.  This results in a disk with a Mach number of 7.7 at the outer boundary, and 24.5 at the inner boundary.

Because this test is axisymmetric (and therefore effectively 1D), it is not very important that the zones move with the flow. However, for demonstration purposes a passive scalar has been added to the initial conditions: $X=\theta(r {\rm cos}(\phi))$ at $t=0$.  This passive scalar is plotted in Figure \ref{fig:kepler1} at time $t=2\pi$, after the flow has had time to shear it into a spiral.

Error is computed identically to the vortex problem. This is computed at a time $t=\pi$, after a half-orbit at the outer boundary, and about 16 orbits at the inner boundary. Truncation error generates transient waves which propagate radially, bouncing between the two boundaries. The $L_1$ error is computed before these transient waves have fully dissipated. Convergence is very close to second-order (Figure \ref{fig:kepler2}). For 512 radial zones, the $L_1$ error is at the $10^{-5}$ level.

Because this problem is supersonic, most of the energy is kinetic, meaning that small relative errors in the energy density can lead to large errors in the pressure. For this reason, it is important to evolve the modified energy density, with $\Omega_E(r)$ chosen to be Keplerian.  Tests in which $\Omega_E(r)$ was set to zero generated very large errors near the inner boundary.

\subsubsection{Cylindrical Kelvin-Helmholtz Instability}
\label{sec:kh}

\begin{figure}
\epsscale{1.0}
\plotone{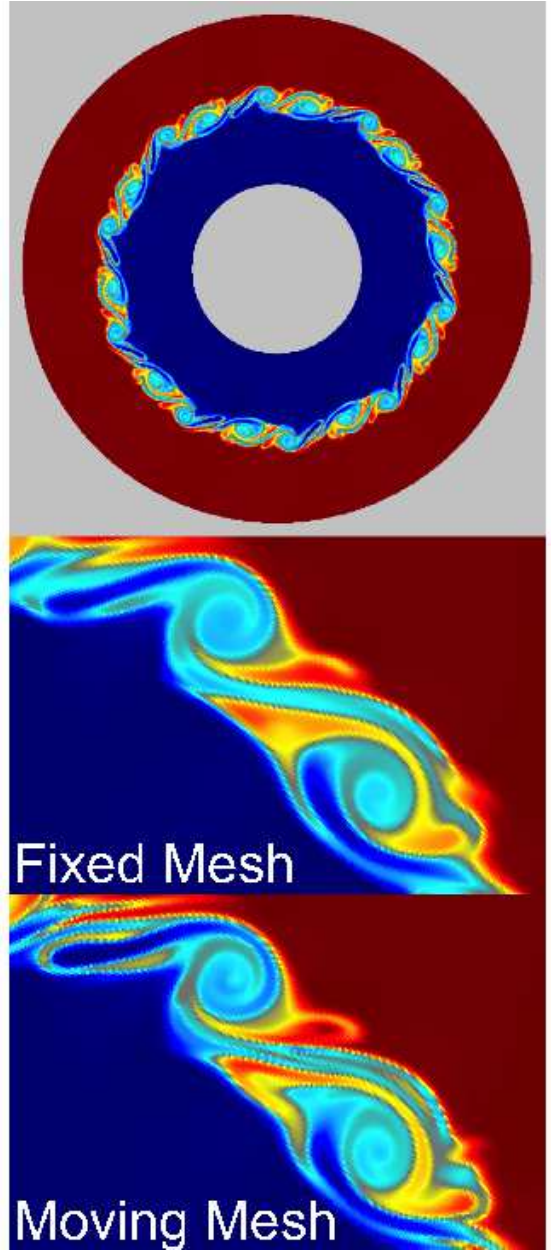}
\caption{ Density at $t = 2.0$ in the cylindrical Kelvin-Helmholtz test.  The upper panel shows the global solution, and the lower two panels are a zoom-in on two of the eddies, showing the difference when the mesh is fixed vs. allowing the mesh to move.  The colormap is the same as in Figure \ref{fig:isentropic1}, but with the density ranging from 1 to 2.
\label{fig:kh} }
\end{figure}

Flows unstable to Kelvin Helmholtz are traditionally tested on cartesian grids.  However, Kelvin-Helmholtz instability can occur in a rotational flow, as shown by the following example.  Radial zones are uniformly distributed from $0.5<r<1.5$.  The background flow is given by a step function across $r=1$:

if $r<1$:
\begin{equation}
\rho = 1,~~
\omega = 2,~~
P = 4+2r^2.
\end{equation}

Otherwise,
\begin{equation}
\rho = 2,~~
\omega = 1,~~
P = 5+r^2.
\end{equation}

The perturbation is introduced in the radial velocity:

\begin{equation}
v_r = v_0 cos(10\phi) e^{- \frac12 (r-1)^2/\sigma_0^2}
\end{equation}
where $v_0=0.02$ and $\sigma_0 = 0.1$.  Figure \ref{fig:kh} plots the density at time $t = 2.0$, showing that the instability is fully nonlinear at this time.  The lower panels show the difference when mesh motion is turned on and off.  Sharper features are present in the version in which the mesh is moved, though care must be taken not to over-interpret how well these features are captured \citep{2016MNRAS.455.4274L}.

\subsection{Viscosity}
\label{sec:visctest}
\subsubsection{Cartesian Shear Flow with Viscosity}

\begin{figure}
\epsscale{1.0}
\plotone{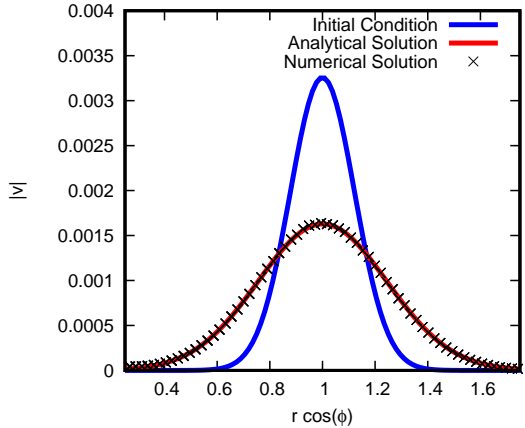}
\caption{ Cartesian shear flow test, comparing DISCO's solution to the analytical Green's function solution given by (\ref{eqn:greens}).  All viscous terms are tested here, as the flow is cartesian, and not aligned with the cylindrical grid.
\label{fig:cartflow} }
\end{figure}

In order to test DISCO's implementation of viscosity, it is necessary to perform a noncircular viscosity test. This is so that every term in the viscous equations is used. Fortunately, this is as simple as setting up a cartesian test problem and putting it on the cylindrical grid.  Of course, the cylindrical grid is not ideal for this test, but the purpose is to make sure all of the terms (\ref{eqn:viscflux}) are implemented correctly, not to test accuracy or convergence.

In cartesian coordinates, if we one designs a flow with uniform density and pressure and with $\vec v = v(x) \hat y$, the Navier-Stokes equations reduce to:

\begin{equation}
\dot v = \nu v'',
\end{equation}
where $\nu$ is viscosity. This has the well-known Green's function solution:

\begin{equation}
v_y = {v_0 \over \sqrt{ 4 \pi \nu t}} exp\{ {-(x-x_0)^2 \over 4 \nu t} \}
\label{eqn:greens}
\end{equation}

This solution is evolved using the following parameters: $\rho=1$, $P=1$, $x_0=1$, $v_0=0.001$, and $\nu=0.03$.

Initial conditions are given by the solution at time t=0.5 and this is evolved until time $t=1.0$.  64 radial zones are uniformly distributed from $0<r<2$.  Figure \ref{fig:cartflow} plots the solution at this time as a function of the x coordinate ($r {\rm cos}(\phi)$) of each zone.  Although the code is not designed for problems so misaligned with the grid, the analytic solution is recovered.

Additionally, the code was run with various terms in equations (\ref{eqn:viscflux}) set to zero, to check the importance of each term.  The test has also been run with various choices of $\Omega_E(r)$, to test that the solution is independent of this choice.  All terms are necessary to capture the solution to this level of accuracy.  In other words, this test ensures that the viscosity is implemented correctly.

\subsubsection{Supersonic Keplerian Spreading Test}

\begin{figure}
\epsscale{1.0}
\plotone{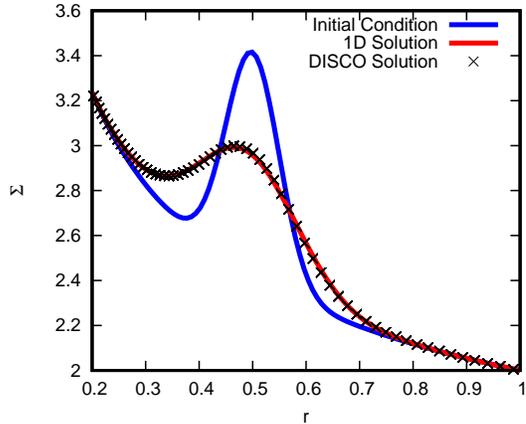}
\caption{ Supersonic Keplerian spreading test.  DISCO's solution is compared with the output of a 1D code which integrates the 1D evolution equation for the surface density (\ref{eqn:1ddisk}).
\label{fig:spread} }
\end{figure}

A more complex viscosity test is attempted, relevant to accretion disks.  A point mass is concentrated at $r=0$ and the domain extends from $0.2<r<1$.  Orbital velocity is Keplerian, $\omega(r) = r^{-3/2}$, and pressure is set to a constant, $P = 0.003$.  A uniform viscosity $\nu = 10^{-5}$ ensures an accretion flow given by $v_r = -\frac32 \nu / r$ (and the initial conditions assume this radial velocity).  The density is given by the following:

\begin{equation}
\Sigma(r) = 1 + 1/\sqrt{r} + e^{-200(r-.5)^2}.
\end{equation}

Note that ``surface density" $\Sigma$ is replacing ``density" $\rho$; this is a cosmetic change which reflects the fact that the code is integrating 1D and 2D (vertically integrated) disk equations.

The first two terms are steady-state solutions to the evolution equations for $\Sigma$, but the final term is a density bump which should be smeared out by viscosity.  In this test, a nearly isothermal equation of state is assumed ($\gamma=1.001$), in order that viscous heating does not cause violations of the thin disk assumptions.

This gives a disk orbiting at about Mach 45 in the vicinity of $r=0.5$.  A very high Mach number is chosen because very cold disks are assumed when deriving the 1D diffusion equation for the surface density:

\begin{equation}
\dot \Sigma = {3 \over r} ( \sqrt{r} ( \sqrt{r} \Sigma \nu )' )'
\label{eqn:1ddisk}
\end{equation}

Similar to the previous test, analytic Green's function solutions exist to this equation.  However, it is also straightforward to write a 1D code to integrate this PDE forward in time, to compare with DISCO's solution.  For the initial conditions stated above for the surface density, a solution was obtained at t = 100 by integrating this PDE.

This is plotted in Figure \ref{fig:spread}, compared with DISCO's solution using 64 radial zones. The solutions do not match identically, but this is likely due to thin-disk assumptions made in deriving the 1D equation.  Errors are at the percent-level.

\subsection{Disk-Planet Interactions}

\subsubsection{Low-Mass Planet}

\begin{figure}
\epsscale{1.0}
\plotone{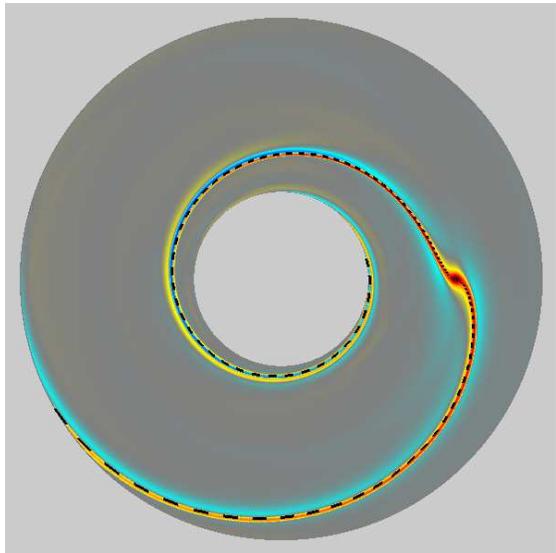}
\caption{ Density for the Earth-mass planet in a Keplerian disk (after ten orbits, with 256 radial zones).  The colormap is the same as in Figure \ref{fig:isentropic1}, but with density ranging between 0.97 and 1.03.  Dashed curve is the analytical formula for the spiral wave given in equation (\ref{eqn:spiral}).
\label{fig:planet1} }
\end{figure}

\begin{figure}
\epsscale{1.0}
\plotone{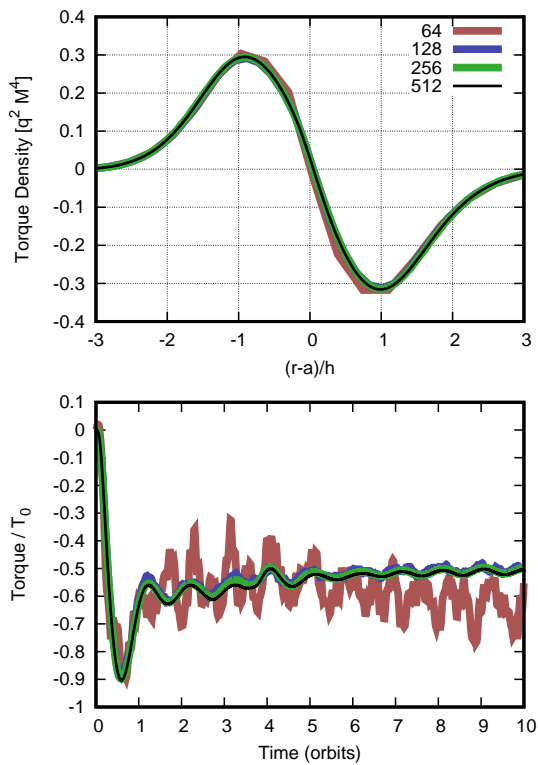}
\caption{ Torque felt by the Earth-mass planet.  The upper panel plots torque density after ten orbits, showing that $128$ radial zones is sufficient to capture this function.  This is mirrored in the lower panel, which shows total torque as a function of time, at various resolutions.
\label{fig:planet2} }
\end{figure}

\begin{figure}
\epsscale{1.0}
\plotone{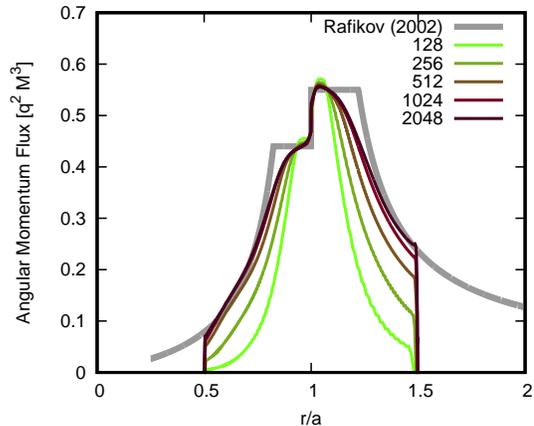}
\caption{ Angular momentum flux carried by the planetary wake, for a planet with $q = 10^{-5}$, corresponding to about three times Earth's mass.  The nonlinear propagation of the wave is compared with the semi-analytic model of \cite{2002ApJ...569..997R}.  The wave shocks a few scale heights away from the planet, outside of which the angular momentum flux drops significantly.  This weak nonlinear effect is much more difficult to capture than linear effects like the torque density.
\label{fig:planet3} }
\end{figure}

It is important to be able to capture planet-disk interactions in an idealized context.  This test explores the linear interaction between an Earth-mass planet and a supersonic Keplerian disk whose Mach number is $\mathcal{M}=20$ at the orbital radius. The initial conditions are not meant to mimic a protoplanetary disk, rather they are meant to produce an idealized environment in which it is easy to test the code.  A small domain is used, with $0.5<r<1.5$ and the planet is located at radius $a=1$.  Orbital velocity is Keplerian ($\omega(r) = r^{-3/2}$), and surface density and pressure are uniform ($\Sigma=1$,  $P=0.0025$).

Here, to simplify the problem, an isothermal equation of state is employed:

\begin{equation}
P = \Sigma / \mathcal{M}^2,
\end{equation}
with boundary conditions fixed at the initial conditions. So far, this is just a supersonic Keplerian disk, similar to the test in section \ref{sec:kepler} (except with an isothermal equation of state). The only additional ingredient is a point mass orbiting at $a=1$.

The planet mass is given by the Earth-to-Sun mass ratio:

\begin{equation}
q = 3 \times 10^{-6} = 0.024 q_{NL}
\end{equation}
where $q_{NL}$ is defined according to the thermal mass threshold for nonlinearity ($q_{NL} = \mathcal{M}^{-3} = 1.25 \times 10^{-4}$).  The planet's gravity is given a smoothing length $\epsilon = 0.5 h = 0.025$.

The simplest analytic comparison is given by the spiral wave produced by the planet. It is straightforward to calculate the shape of this wave from linear theory \citep[e.g.][]{2002MNRAS.330..950O}:

\begin{equation}
\phi(r)= \phi_p + \text{sign}(r-a)(3-2\sqrt{a/r}-r/a) \mathcal{M}.
\label{eqn:spiral}
\end{equation}

The spiral wave is established in the first orbit ($t = 2\pi$).  Figure \ref{fig:planet1} shows the density after $10$ orbits, also plotting the analytical prediction (\ref{eqn:spiral}) for the spiral.

Time-averaged torque density exerted by the planet is shown in Figure \ref{fig:planet2} for various resolutions, showing convergence of the torque.  The total torque is affected by the nearby boundaries, but it reaches its converged value with only $128$ radial zones.

Nonlinear propagation of the spiral wave is a much more subtle and difficult behavior to capture \citep{2011ApJ...741...57D}.  This can be measured by the angular momentum flux (AMF) emanating from the planet.  This AMF has two components, a gravitational component due to the excitation of the wave, and a ``wave" component due to the wave's propagation and dissipation.

\begin{equation}
\Phi_p = \Phi_{\rm grav} + \Phi_{\rm wave}
\end{equation}
\begin{equation}
\Phi_{\rm grav} = \left\{ \begin{array}{rl}
  \int_r^{\infty} {dT \over dr} dr & ~ r > a \\
  \\
  -\int_0^{r} {dT \over dr} dr     & ~ r < a
       \end{array} \right.
\end{equation}
\begin{equation}
\Phi_{\rm wave} = \int r d\phi \Sigma v_r r^2 (\omega - \omega_K),
\end{equation}
where $\omega_K$ is the background Keplerian orbital frequency.  \cite{2001ApJ...552..793G} produced semi-analytical formulas for wave propagation and dissipation, which were later generalized to a global disk \citep{2002ApJ...572..566R}.  These scalings are plotted in Figure \ref{fig:planet3} compared with the measured planetary AMF for a planet with $q = 10^{-5}$ (about three times Earth's mass).  Convergence is harder to establish for this test than for the linear wave.  $1024$ zones ($46$ per scale height) are needed for reasonably accurate measurement of the AMF.

\subsubsection{Viscous Disk with a Gap-Opening Planet}

\begin{figure}
\epsscale{1.0}
\plotone{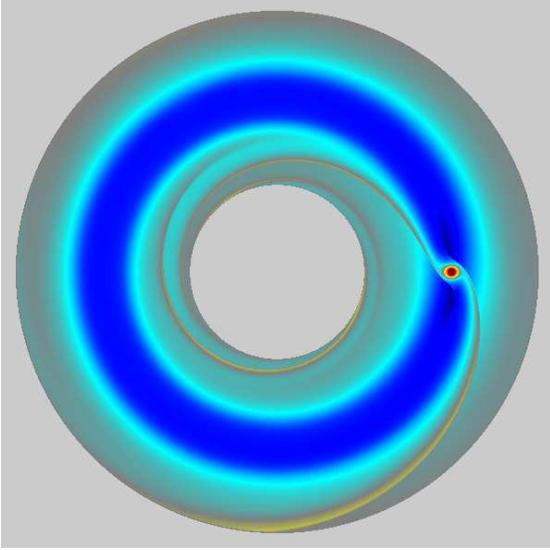}
\caption{ Jupiter-mass planet in a Viscous Disk after $1000$ orbits, $\nu = 10^{-5}$.  Because the planet mass is much larger than in previous tests, the wave produces shocks strong enough to open a deep gap.  Colormap is the same as in Figure \ref{fig:isentropic1}, but with density in logscale ranging from $10^{-2}$ to $10^2$.
\label{fig:jupiter1} }
\end{figure}

\begin{figure}
\epsscale{1.0}
\plotone{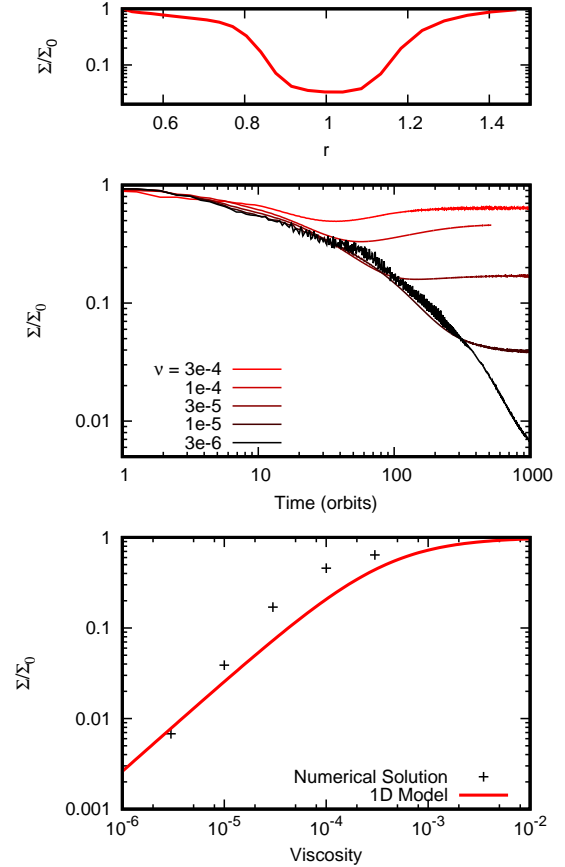}
\caption{ Gap depth for Jupiter in a viscous disk.  The top panel shows azimuthally averaged surface density as a function of radius for the $\nu = 10^{-5}$ case.  The center panel shows gap depth as a function of time for various viscosities.  The lower panel shows the final gap depth (after $1000$ orbits) as a function of viscosity, comparing with the analytical 1D model for the gap depth (\ref{eqn:1ddepth}).
\label{fig:jupiter2} }
\end{figure}

\begin{figure*}
\epsscale{1.0}
\plotone{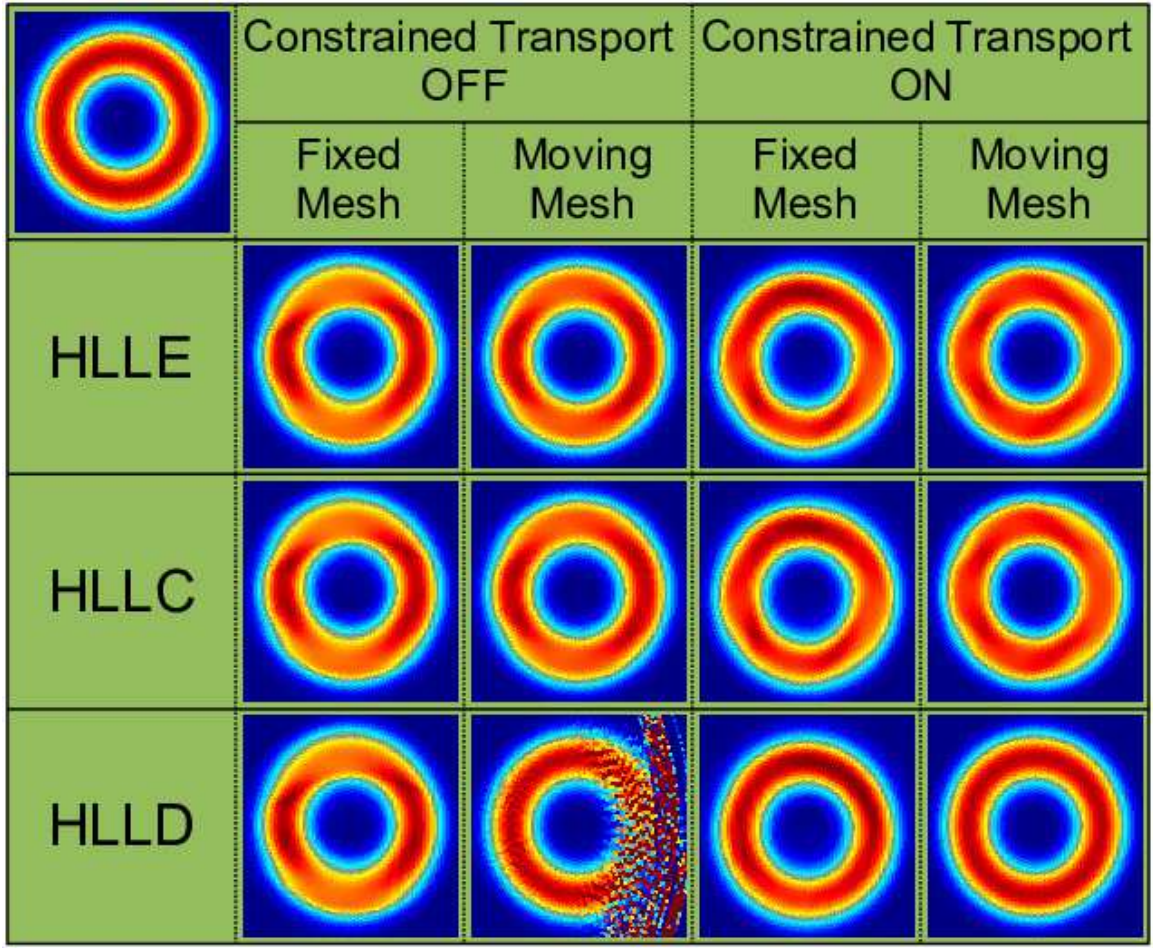}
\caption{ Magnetic field loop after advecting one orbit, using various methods.  When constrained transport is turned off, $B_r$ is evolved differently from $B_\phi$, resulting in an asymmetric loop, and potentially resulting in unstable evolution.  With CT turned on, and using the HLLD Riemann solver on the moving mesh, the loop can be maintained to machine precision.  Colormap is the same as in Figure \ref{fig:isentropic1}, but magnetic energy is plotted, ranging from zero to $5.25 \times 10^{-9}$.
\label{fig:floop} }
\end{figure*}

Larger planets produce stronger shocks, which dissipate the angular momentum in the spiral wave, depositing torque in the disk, which causes an evacuation of an annulus in the vicinity of the planet's orbit.  This low-density annulus is known as a ``gap", and it is well-known that a Jupiter-mass planet can open a gap in a disk with moderate viscosity.

The scaling of gap depth with viscosity has been determined in several numerical studies \citep{2013ApJ...769...41D, 2014ApJ...782...88F, 2015MNRAS.448..994K, 2015ApJ...807L..11D}:

\begin{equation}
{\Sigma_{\rm gap} \over \Sigma_0} = { 1 \over 1 + f_0 K(q) / 3 \pi },
\label{eqn:1ddepth}
\end{equation}
where $f_0 = 0.45$ and

\begin{equation}
K(q) \equiv q^2 \mathcal{M}^5 / \alpha.
\end{equation}

This scaling will be measured using the same setup as the previous test, but with $q = 10^{-3}$ (Jupiter-to-sun mass ratio).  Figure \ref{fig:jupiter1} shows the density at $10^3$ orbits with $\nu = 10^{-5}$.  Figure \ref{fig:jupiter2} measures the gap depth as a function of time, and compares the steady-state gap depth with the prediction from empirical calculations and 1D models (\ref{eqn:1ddepth}).

\cite{2013ApJ...769...41D} used this test as a means to build a working definition of ``numerical viscosity" in DISCO.  The same test is performed with zero viscosity, and using a number of different numerical resolutions, and the gap depth was measured as a function of resolution.  This was compared to the gap depth as a function of viscosity, to build a definition of ``effective viscosity" of the numerical scheme.  In that study, it was found that 

\begin{equation}
\alpha_{\rm num} = 2.5 \times 10^{-3} (\Delta r / h)^2.
\end{equation}

\subsection{Magnetohydrodynamics}

It will be important to test the accuracy and stability of the CT scheme described in section \ref{sec:mhd}.  Many works testing numerical MHD studies look at the magnitude of div B for given test problems, as a guide to code performance.  Yet, the magnitude of div B can be small and still affect the long-term behavior of the solution.

The approach taken here to diagnosing div B performance is to measure the indirect impact of the divergence on the final solution.  For example, one can advect a field loop until ${\nabla \cdot B}$ errors produce inaccurate evolution.  Then one can demonstrate whether enforcing the div B constraint resolves the issue.  This approach leads to a much greater confidence that attempts to eliminate magnetic monopoles are accomplishing something.

\subsubsection{Orbital Advection of a Field Loop}

The test which is typically most sensitive to div B errors is the advection of a magnetic field loop.  A modified version of this test is presented here to test DISCO's CT algorithm in 2D.  The initial setup (in the domain $0 < r < 0.5$) is given as follows:

\begin{equation}
\rho = 1, \omega = 1, P = 0.01 + 0.5 r^2.
\end{equation}

The magnetic field is given by:

\begin{equation}
B = \left\{ \begin{array}{rl}
  B_0 \text{sin}^2( \pi \tilde r / R ) \sqrt{2 \tilde r / R} & ~ \tilde r < R \\
  0 & ~ \tilde r > R
       \end{array} \right.
\end{equation}
where $\tilde r$ is the distance from a point centered at $r = 0.25$, $\phi = 0$.  $R = 0.15$ sets the radius of the loop.  This magnetic field loop has magnetic tension and pressure, which can be balanced with the following adjustment to the gas pressure:

\begin{eqnarray}
P \rightarrow P - B_0^2 ( (\tilde r/R) \text{sin}^4( \pi \tilde r / R ) + \\
{ 12 \pi \tilde r / R - 8 \text{sin}( 2 \pi \tilde r / R ) + \text{sin}( 4 \pi \tilde r /R ) \over 16 \pi } ) \nonumber
\end{eqnarray}

However, this adjustment is largely unnecessary, as the magnetic field strength is very weak ($B_0 = 10^{-4}$).  Stronger magnetic fields are possible to use, but they are less susceptible to numerical disruption, as magnetic tension and pressure provide restoring forces to the loop.

Figure \ref{fig:floop} shows the field loop after a single orbit advected using $N_r = 128$ radial zones, using 12 different combinations of schemes (with different options for the Riemann solver, mesh motion, and CT).  HLLD shows a significant improvement over the other schemes, reducing the diffusion of the field loop.  However, the loop is more susceptible to div B instability when using this less diffusive Riemann solver.  More diffusive methods mitigate immediate catastrophe by diffusing out div B errors, but they do not eliminate the problem.  Employing CT and HLLD, the loop is preserved to high precision, and with the mesh motion the advection is solved to machine precision (ignoring errors in the pressure balance, which are small).

\subsubsection{Spinning 3D Loop}

\begin{figure}
\epsscale{1.0}
\plotone{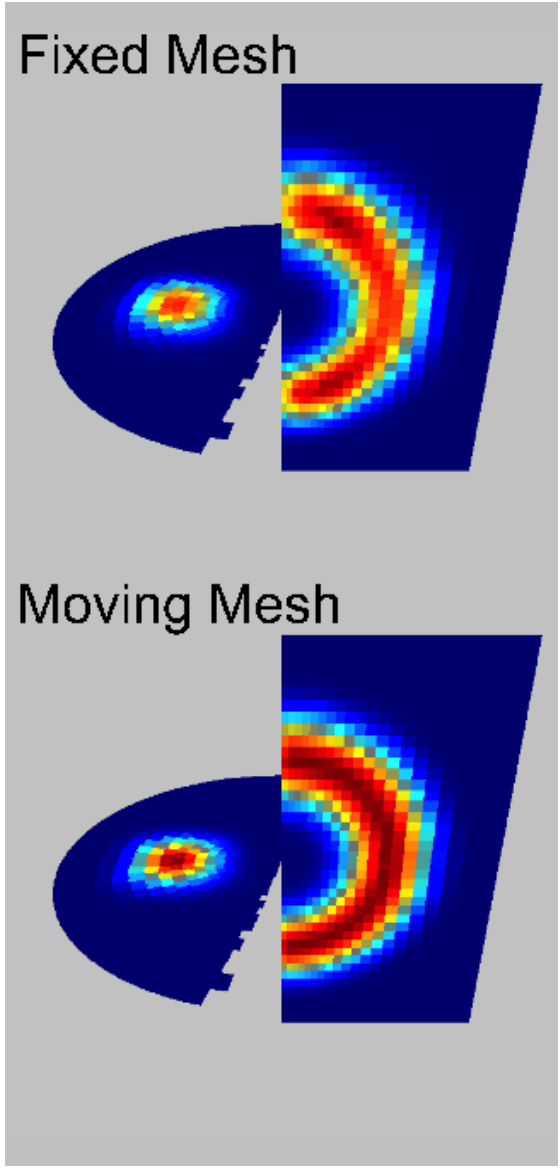}
\caption{ 3D spinning field loop test after a single orbit, similar to the orbital advection test, but in three dimensions.  A fixed mesh solution is compared with a moving mesh solution, showing the diffusion of the spinning loop that results from advection errors.  Colormap is the same as in Figure \ref{fig:isentropic1}, but magnetic energy is plotted, ranging from $0$ to $5.25 \times 10^{-21}$.
\label{fig:sploop} }
\end{figure}

Testing the CT algorithm in 3D necessitates a 3D magnetic field configuration, with nonzero components of $B_r$, $B_{\phi}$ and $B_z$.  The spinning 3D loop initially has field components $B_x$ and $B_z$, and this loop rotates about the z-axis rigidly.  Initial conditions are specified as:

\begin{equation}
\rho = 1.0, ~\omega = 1.0, ~P = 1.0 + .5 r^2.
\end{equation}
\begin{equation}
B_x = - B_1 z/\tilde r, ~B_z = B_1 x/\tilde r
\end{equation}
where $\tilde r$ is a radial coordinate in the x-z plane:
\begin{equation}
\tilde r = \sqrt{x^2+z^2}
\end{equation}
\begin{equation}
B_1 = \left\{ \begin{array}{rl}
  B_0 \text{sin}^2(\pi \tilde r/R) \text{cos}(\pi y/R) \sqrt{2 \tilde r \over R} & ~ \tilde r < R, |y| < .5 R \\
  0 & ~ \text{otherwise}
       \end{array} \right.
\end{equation}
and this magnetic field is chosen to be very weak:

\begin{equation}
B_0 = 10^{-10}
\end{equation}

Figure \ref{fig:sploop} shows the spinning 3D loop using only $16$ radial zones ($32$ vertical zones) after a single orbit.  The upper panel uses a fixed mesh, while the lower panel uses a moving mesh.  Moving the mesh allows for accurate preservation of this loop.

\subsubsection{MHD Flywheel}

\begin{figure}
\epsscale{1.0}
\plotone{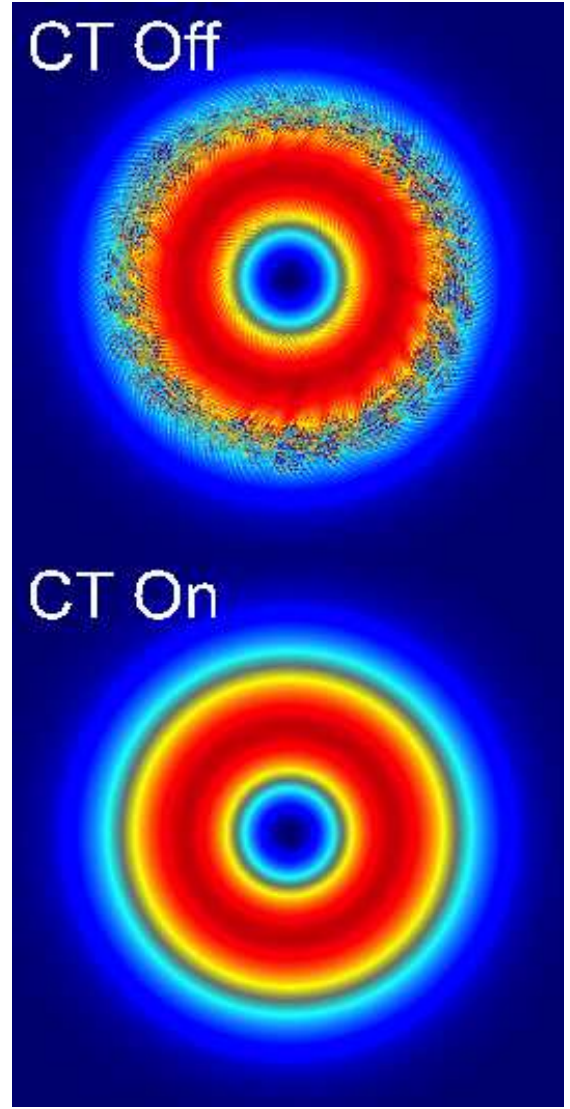}
\caption{ 2D MHD Flywheel test at $t=5$.  The upper panel has constrained transport turned off, while the lower panel has constrained transport on.  CT is necessary for numerical stability on this test; the grid-scale high-amplitude noise on the top panel is due to numerical instability.  Colormap is the same as in Figure \ref{fig:isentropic1}, but magnetic energy is plotted, ranging from 0 to $4.5 \times 10^{-3}$.
\label{fig:fly1} }
\end{figure}

\begin{figure}
\epsscale{1.0}
\plotone{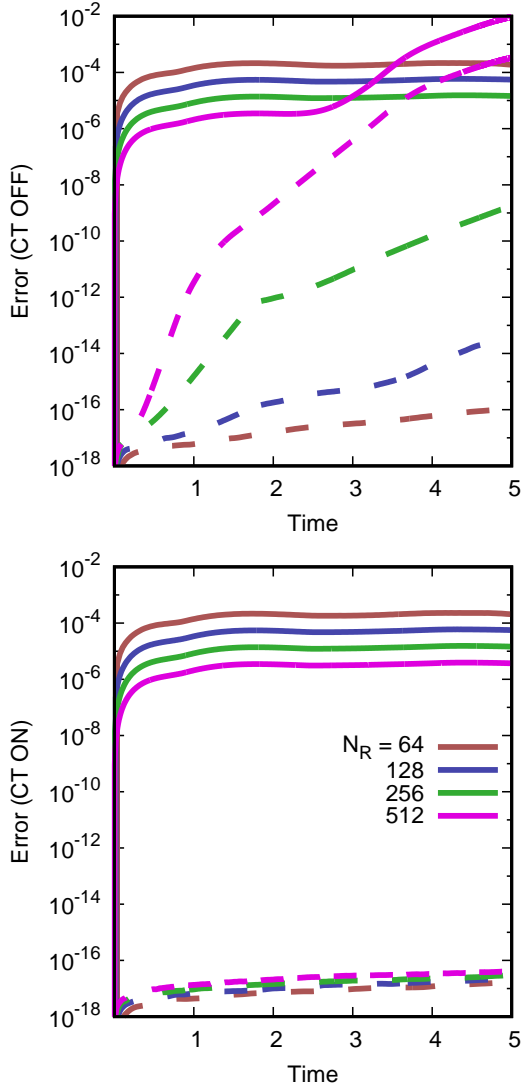}
\caption{ $L_1$ errors for the MHD flywheel test.  Solid curves indicate error in the density (equation \ref{eqn:l1rho}), and dashed curves signify error in the radial component of the magnetic field (equation \ref{eqn:l1B}).  The upper panel has uncontrolled divergence errors, whereas the lower panel avoids this numerical instability by employing constrained transport.
\label{fig:fly2} }
\end{figure}

The ``MHD Flywheel" test is a new test presented here which illustrates the numerical instability associated with div B errors.  The setup involves a stationary orbital MHD configuration, in which centrifugal force is balanced by magnetic tension, and magnetic pressure is balanced by gas pressure:

\begin{equation}
\rho = 1,~~ v_r = 0,~~ \omega = \Omega_0 e^{-\frac12 r^2/R^2},
\end{equation}

\begin{equation}
\vec B = \sqrt{\rho} v_{\phi} \hat \phi,~~ P = P_0 - \frac12 B^2.
\end{equation}

This gives a stationary solution in 3D (though this test is only performed in 2D).  Note that since velocity is parallel to $\vec B$, the total azimuthal flux of $\vec B$ is zero, which makes this test succeptible to div B violations if CT is not employed.  Constants are chosen so that

\begin{equation}
P_0 = 1.1 (\frac12 e^{-1}) \rho \Omega_0^2 R^2,
\end{equation}
\begin{equation}
\Omega_0 = 1.5,~~ R = 0.1.
\end{equation}

In 2D, this stationary solution is stable, so the goal is to preserve the initial conditions as precisely as possible.  Figure \ref{fig:fly1} shows the solution at $t=5$ using high resolution ($N_r = 512$) and toggling CT on and off.  While CT is turned off, div B errors cause a numerically unstable solution.  Implementing constrained transport eliminates these errors.

This is shown quantitatively in Figure \ref{fig:fly2}.  Here, errors in the solution are plotted.  These are measured in both the density and the radial component of the magnetic field:

\begin{equation}
L_1^{\rho} = { \int | \rho - 1.0 | dV \over \int dV }.
\label{eqn:l1rho}
\end{equation}

\begin{equation}
L_1^{B} = { \int | B_r | dV \over \int dV }.
\label{eqn:l1B}
\end{equation}

Both errors are plotted in Figure \ref{fig:fly2}, with CT turned off on the upper panel, and implemented in the lower panel.  Without controlling div B, it is clear that errors in $B_r$ grow exponentially from machine round-off.  Moreover, the growth rate appears to be proportional to the resolution, so that improving the resolution gives greater errors.  Finally, when these errors grow from machine round-off to be non-negligible contributions to the hydro evolution, they eventually affect the errors in density (which would otherwise converge at second-order).  By implementing constrained transport, the errors in $B_r$ remain very small.

\subsubsection{Cylindrical MHD Explosion}

\begin{figure}
\epsscale{1.0}
\plotone{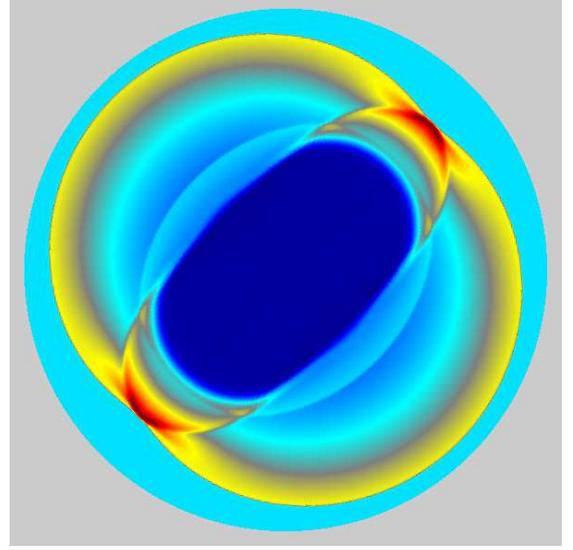}
\caption{ MHD explosion test at $t=0.2$, using $256$ radial zones.  Colormap is the same as in Figure \ref{fig:isentropic1}, but density ranges from $0.1$ to $2.7$.  The origin is not excised from the grid.  DISCO is able to capture all of the nontrivial shock structures with or without CT.
\label{fig:bexp} }
\end{figure}

A commonly used setup which is useful for comparisons with other codes, this test puts an explosion into an initially uniform magnetic field, pushing the field lines sideways and generating a nontrivial shock structure.

The computational domain extends from $0 < r < 0.55$.  Density is uniform $\rho = 1.0$, and all velocities are initially zero.  A magnetic field with strength $B_0 = 1.0$ is pointed along the $45^{\circ}$ diagonal (though the direction of the field is irrelevant, as the computational domain is cylindrical).

The pressure is set to $P = 0.1$ everywhere except within $r < 0.1$, where $P = 10$.  Figure \ref{fig:bexp} shows the explosion at time $t = 0.2$ using $N_r = 256$ radial zones (roughly equivalent to a cartesian box with $512$ zones across).  DISCO's performance on this test can be compared e.g. with Figure 28 of \cite{2008ApJS..178..137S}, or Figure 4 of \cite{2011MNRAS.418.1392P}.

Note that the origin is kept on the grid, despite the coordinate singularity at $r=0$.  In practice, the numerical solution to this test did not significantly depend on whether CT was employed.

\subsubsection{MHD Rotor}

\begin{figure}
\epsscale{1.0}
\plotone{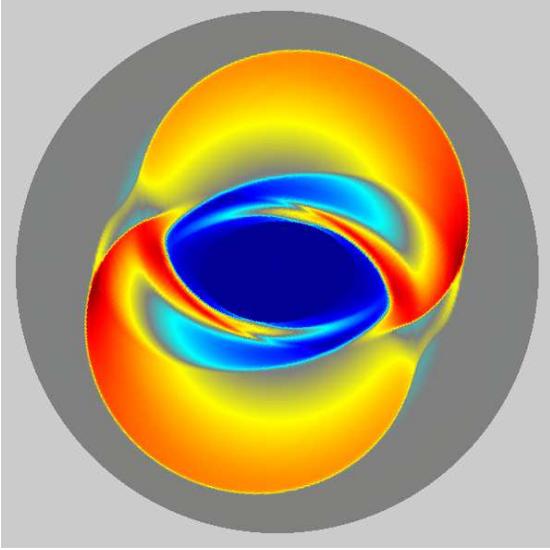}
\caption{ MHD rotor test at $t=0.15$, using $256$ radial zones.  DISCO captures all of the details of the rotor, and is consistent with other MHD codes on this test, whether or not CT is employed.  Colormap is the same as in Figure \ref{fig:isentropic1}, but gas pressure is plotted, ranging from $0$ to $2$.  The origin is not excised from the grid.
\label{fig:rotor} }
\end{figure}

This test is more challenging than the MHD explosion, and it tests the code's propagation of strong torsional Alfv\'en waves.  A rotational flow at the center winds up an initially uniform magnetic field:

\begin{equation}
\omega = \left\{ \begin{array}{rl}
  v_0/R_0 f(r) & ~ r < R_0 \\
  v_0/r f(r)   & ~ R_0 < r < R_1 \\
  0            & ~ r > R_1
       \end{array} \right.
\end{equation}
where $f(r)$ is a tapering function:

\begin{equation}
f(r) = \left\{ \begin{array}{rl}
  1                 & ~ r < R_0 \\
  (R_1-r)/(R_1-R_0) & ~ R_0 < r < R_1 \\
  0                 & ~ r > R_1
       \end{array} \right.
\end{equation}

Density, pressure, and radial velocity are given by

\begin{equation}
\rho = 1 + 9 f(r), ~~ P = 1, ~~ v_r = 0
\end{equation}
and the magnetic field is uniform in the x direction:

\begin{equation}
\vec B = (5/\sqrt{4 \pi}) \hat x.
\end{equation}

Another important point to note is that unlike the rest of these test problems, which have either used $\gamma = 5/3$ or an isothermal equation of state, this test uses a lower adiabatic index:

\begin{equation}
\gamma = 1.4
\end{equation}

Figure \ref{fig:rotor} shows the pressure at $t = 0.15$, which can be compared with Figure 2 of \cite{1999JCoPh.149..270B} or Figure 25 of \cite{2008ApJS..178..137S}.

Again, the origin at $r=0$ is resolved, and like the MHD explosion, the solution did not depend significantly on whether CT was employed.

\subsubsection{3D MHD Explosion}

\begin{figure}
\epsscale{1.0}
\plotone{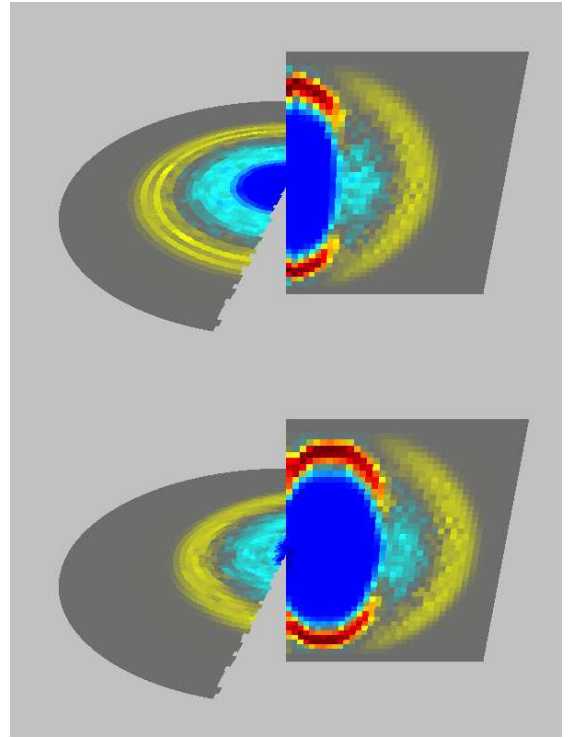}
\caption{ 3D MHD Explosion at time $t = 0.1$.  Colormap is the same as in Figure \ref{fig:isentropic1}, but with density taking on values between $0$ and $2$.  This test uses $35$ radial zones and $45$ vertical zones.  The lower panel is the same test, but offset from the origin by $x_{\rm off} = 0.1$.  This test is not performed in other studies, but confidence in the scheme is gained in that the code attains the same solution when the explosion is offset from the origin, and the test is no longer axisymmetric.  Note in the lower panel that the pressure jump is initiated across the origin in the midplane.
\label{fig:bx3d} }
\end{figure}

In order to test DISCO's performance on a 3D MHD test problem with nontrivial shocks, a 3D explosion test is performed.  Initial conditions are given by

\begin{equation}
\rho = 1,~~\vec B = 2 \hat z,
\end{equation}

\begin{equation}
P = \left\{ \begin{array}{rl}
  10 & ~ r < 0.1 \\
  0.1   & ~ r > 0.1 \\
       \end{array} \right.
\end{equation}

All velocities are initially zero.  Figure \ref{fig:bx3d} shows the 3D explosion at time $t = 0.2$.  A second test was performed where the explosion is offset from the origin by an amount $x_0 = 0.1$, in order to create a truly three-dimensional (non-axisymmetric) test; this is plotted in the lower panel of Figure \ref{fig:bx3d}.  This test is not performed in other studies, so there do not exist other examples to compare with, but the test demonstrates basic capturing of a 3D MHD shock structure, and that all geometric terms are correctly implemented, as the offset test has an equivalent solution to the centered test.  Additionally, this test shows that DISCO is capable of evolving shocks propagating along the coordinate axis, and shocks colliding with the coordinate singularity at $r=0$, especially in the offset case, where the pressure jump is initiated across $r=0$ in the equatorial plane.

\subsubsection{Magnetorotational Instability: Linear Growth}

\begin{figure}
\epsscale{1.0}
\plotone{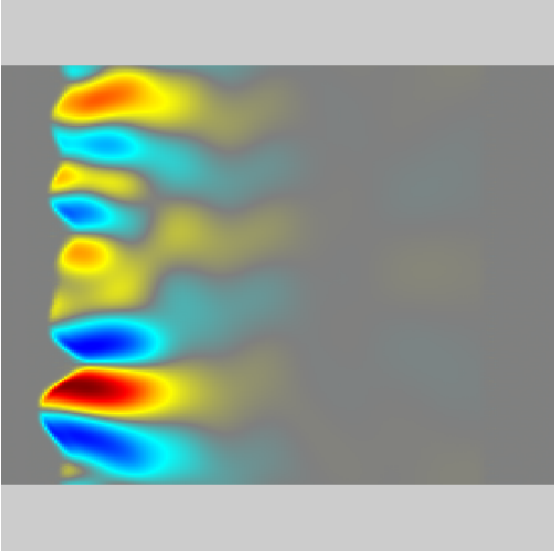}
\caption{ Flock 3D MRI test after eight orbits, with $N_r \times N_z = 512 \times 256$ (x-z plane is shown).  A coherent unstable mode emerges out of white noise initial perturbations \citep[compare with Figure 5 of][]{2010AnA...516A..26F}.  Colormap is the same as in Figure \ref{fig:isentropic1}, but with azimuthal magnetic field ranging from -0.02 to 0.02.
\label{fig:flock1} }
\end{figure}

\begin{figure}
\epsscale{1.0}
\plotone{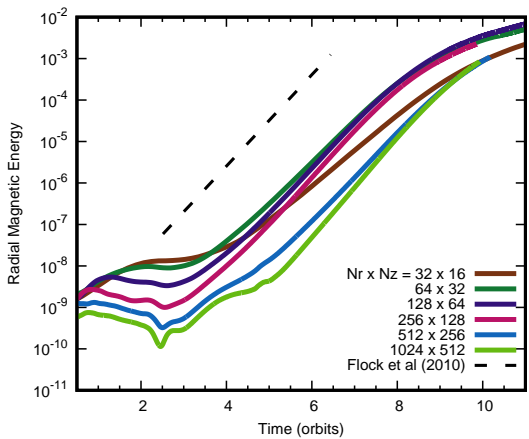}
\caption{ Exponential growth of the MRI in the 3D Flock test.  Dashed curve is the growth rate found by \cite{2010AnA...516A..26F} using the HLLD and Roe solvers using an upwind CT scheme.  This rate is consistent with the analytical MRI growth rate.
\label{fig:flock2} }
\end{figure}

One of the most important astrophysical applications of numerical MHD is the study of the magnetorotational instability \citep[MRI][]{1991ApJ...376..214B, 1995ApJ...440..742H, 1996ApJ...463..656S}.  MRI is thought to be the source of angular momentum transport (and hence accretion) in most astrophysical disks.  Growth of MRI using DISCO will be studied using two test problems.  First, linear growth is explored, by implementing the 3D MRI test of \cite{2010AnA...516A..26F}.  The set-up was described very clearly in that study, so the initial conditions will not be repeated here.

The velocity field is seeded with white noise, and the magnetic field is chosen so that $n=4$ corresponds to the fastest-growing MRI mode.  Figure \ref{fig:flock1} shows the azimuthal component of the magnetic field after eight orbits, to be compared with Figure 5 of \cite{2010AnA...516A..26F}.  Coherent magnetic fields have grown from white noise initial perturbations.  Growth of the MRI is shown in Figure \ref{fig:flock2}, which plots the ``Radial magnetic energy", showing the exponential growth of the instability, compared with the growth rate found by \cite{2010AnA...516A..26F} (see their Figure 3).  At resolutions above $N_z = 32$, or $8$ zones per MRI wavelength, the correct growth rate is recovered.

\subsubsection{Magnetorotational Instability: Nonlinear Saturation}

\begin{figure}
\epsscale{1.0}
\plotone{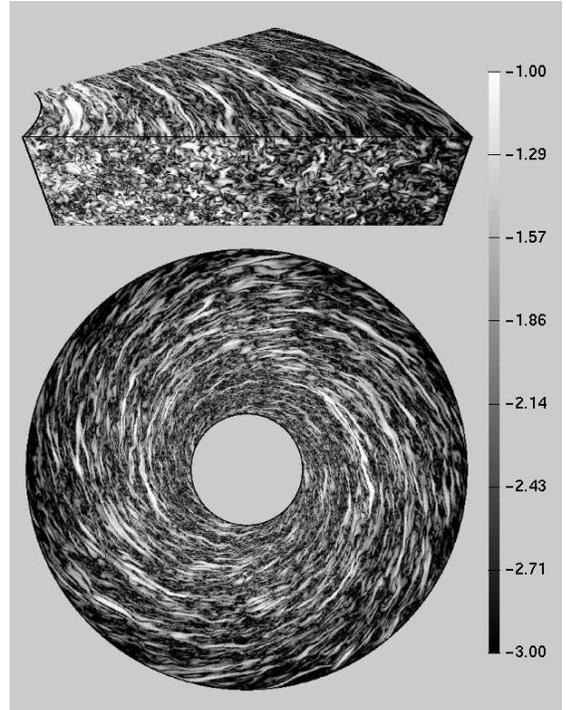}
\caption{ Nonlinear 3D MRI test after 25 orbits.  Logarithm of magnetic energy is plotted.  This test demonstrates DISCO's ability to capture MRI, not just in the linear growth phase, but in the nonlinear, fully turbulent regime.
\label{fig:stone1} }
\end{figure}

\begin{figure}
\epsscale{1.0}
\plotone{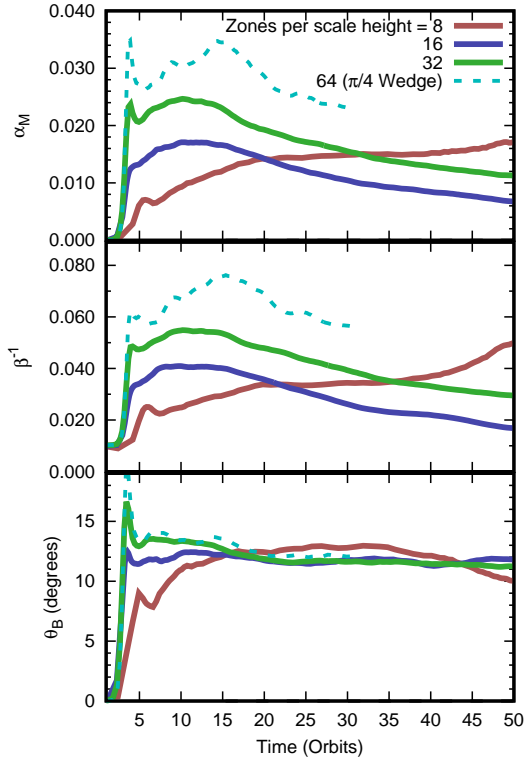}
\caption{ Growth and saturation in the 3D nonlinear MRI test.  The various resolutions shown covered two scale heights, so that $N_r \times N_z = 64 \times 16$, $128 \times 32$, $256 \times 64$, and $512 \times 128$.  Solid curves used numerical grids which spanned $2 \pi$ in azimuth.  The cyan curve is a calculation on a wedge spanning $\pi/4$ in azimuth.  These measurements can be compared with the detailed numerical MRI study of \cite{2012ApJ...749..189S}.  In particular, \cite{2012ApJ...749..189S} found a magnetic tilt angle of $12^{\circ}$, consistent with the angle found in the present study (bottom panel).
\label{fig:stone2} }
\end{figure}

Capturing the correct linear growth rate of MRI is an important benchmark for any 3D MHD code, but it should be recognized that linear growth does not constitute a complete MRI test.  The true test comes in the nonlinear, fully turbulent phase, when the statistical properties of the turbulence could be affected by the numerical scheme.

The nonlinear MRI test of \cite{2012ApJ...749..189S} is implemented here, to determine DISCO's performance on a 3D nonlinear turbulent magneto-rotational flow.  The initial conditions used are given by their ``zero net flux" case.  Figure \ref{fig:stone1} shows a snapshot in time of the magnetic energy, to be compared with Figure 1 of \cite{2012ApJ...749..189S}.  Figure \ref{fig:stone2} plots the quantities $\alpha_M$ and $\beta$, which are given by

\begin{equation}
\alpha_M = - \left< B_r B_\phi \right> / \left< P \right>,
\end{equation}
\begin{equation}
\beta = \left<P\right> / \left<\frac12 B^2\right>.
\end{equation}
where the average is taken over the entire domain.  These can be compared with \cite{2012ApJ...749..189S}, Figure 5.  Finally, the magnetic ``Tilt angle" is measured.  This is defined as

\begin{equation}
\theta_B = \text{sin}^{-1}(\alpha \beta)/2.
\end{equation}

\cite{2012ApJ...749..189S} showed that this tilt angle is robustly $12^{\circ}$ (see their Figure 11 and their Table 3).  In the bottom panel of Figure \ref{fig:stone2}, this tilt angle is plotted, showing consistency with this value.

\subsection{Efficiency and Scaling}
\label{sec:scaling}

\begin{figure}
\epsscale{1.0}
\plotone{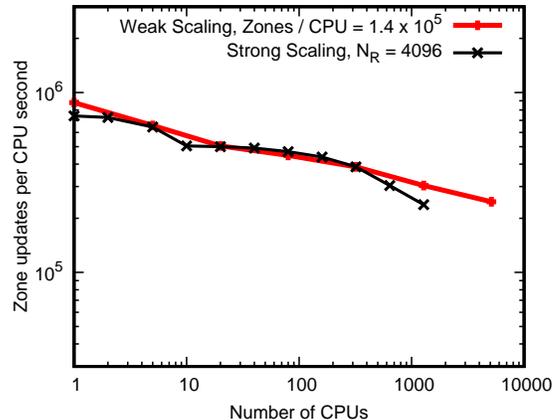}
\caption{ Efficiency and scaling of DISCO for the Keplerian shear flow test (section \ref{sec:kepler}).  When run in parallel, DISCO typically achieves about $5 \times 10^5$ zone updates per CPU second.  Both strong and weak scaling preserve this rate up to thousands of CPUs.
\label{fig:scaling} }
\end{figure}

Finally, a few tests are performed to determine DISCO's performance and scaling to large numbers of processors.  The Keplerian shear flow test (section \ref{sec:kepler}) is performed in both 2D and 3D, on various numbers of processors, and with variable resolution.  Both ``strong scaling" (with fixed resolution) and ``weak scaling" (with fixed number of zones per process) are measured in Figure \ref{fig:scaling}.  Calculations wer performed with 2.8 GHz Intel Xeon E5-2680v2 processors, using the 20-core ``Ivy-Bridge" nodes on NASA's Pleiades supercomputer.  DISCO's peak performance is $9 \times 10^5$ zone updates per CPU second.  In the range between a few nodes and up to thousands of CPUs, typical performance is $5 \times 10^5$ zone updates per CPU second.  A latency is reached when there are as few as two radial zones per CPU in 2D (or two radial and two vertical zones per CPU in 3D).

\section{Discussion}
\label{sec:discussion}	

A moving-mesh technique is presented for numerically integrating both hydro and MHD equations in 3D, with specific application to astrophysical disks, using a new constrained transport technique.  The DISCO code has been made publicly available at \texttt{https://github.com/duffell/Disco} under the GNU general public license.

The numerical scheme has been detailed in this work, including a description of the novel constrained transport technique.  The versatility of the scheme is shown by describing the implementation of many different systems of equations, including the ``bare" Euler equations, viscous hydrodynamics, magnetohydrodynamics, and including additional terms due to the gravitational influence of orbiting bodies.

Many code tests have been performed, demonstrating DISCO's ability to integrate all of these different systems of equations.  DISCO has no trouble accurately capturing shocks, whether they are aligned or misaligned with the orbital motion.  DISCO excels at advecting contact discontinuities and MHD discontinuities with the orbital flow.  DISCO also converges at second-order for smooth flows, and can evolve flows with very high Mach number without accumulating significant errors.

DISCO is ideal for studies of disk-satellite interactions, especially problems which involve a near-equal-mass binary, where both point masses must be resolved on the grid.  It is also an accurate code for calculating MRI, as demonstrated by code tests of both linear stability and nonlinear turbulence.

The constrained transport technique for maintaining zero divergence errors while updating the magnetic fields is unique and applicable to nontrivial and dynamic mesh topologies.  Stable, accurate evolution is possible with this CT method and, as demonstrated in several of the code tests, CT is necessary to prevent inaccurate or unstable evolution.

Certainty in the robustness of the CT method is assured by DISCO's accurate calculation of the magnetorotational instability, capturing both the linear growth rate and the nonlinear turbulent phase in quantitative comparisons with previous numerical studies.

In the future, it will also be possible to use DISCO to integrate the equations of general relativistic magnetohydrodynamics (GRMHD), for application to black hole accretion disks.  Self-gravity is another future improvement which is possible.  Both of these improvements will be presented in a future study.

\acknowledgments

Resources supporting this work were provided by the NASA High-End Computing (HEC) Program through the NASA Advanced Supercomputing (NAS) Division at Ames Research Center.  Some early exploratory calculations employed the Savio computational cluster resource provided by the Berkeley Research Computing program at the University of California, Berkeley (supported by the UC Berkeley Chancellor, Vice Chancellor of Research, and Office of the CIO).  I am grateful to Andrew MacFadyen, Daniel D'Orazio, Brian Farris, Jeffrey Fung, Andrei Gruzinov, Zoltan Haiman, Frederic Masset, Philip Mocz, Diego Mu\~noz, Eliot Quataert, Stephan Rosswog, Geoff Ryan, Jim Stone, and Jonathan Zrake for helpful comments and discussions.  I would also like to thank the anonymous referee for the very helpful review.

\begin{appendix}
\section{Viscosity}

In section \ref{sec:visc_sec}, viscous terms were added to Euler's equations.  Here these terms are derived.  The Navier-Stokes equation is given by the following:

\begin{equation}
\partial_t ( \rho \vec v ) + \nabla \cdot ( \rho \vec v \vec v ) + \vec \nabla P = \nabla \cdot \sigma,
\end{equation}
Where $\sigma$ is the viscous stress tensor:
\begin{equation}
\tensor \sigma = \rho \nu ( \vec \nabla \vec v + \vec \nabla \vec v  + \eta \nabla \cdot v \tensor I )
\end{equation}

The quantity $\eta$ is related to the ratio of bulk viscosity to shear viscosity.  For pure shear flow, the value of $\eta$ should have no impact on the solution.  Initially, the value of $\eta = 0$ is chosen, and will be re-introduced near the end of the derivation.  Additionally, for ease of presentation, the prefactor $\rho \nu$ will be ignored (set to unity) until the last few steps.  Finally, this derivation will be carried out in 2D ($r,\phi$) for simplicity, but the additional terms in the vertical dimension are straightforward to calculate.

The derivatives in the formula for $\sigma$ are covariant derivatives, so in index notation the expression evaluates as:

\begin{equation}
\sigma_{ij} = \nabla_i v_j + \nabla_j v_i = \partial_i v_j + \partial_j v_i - 2 \Gamma^k_{ij} v_k,
\end{equation}
where $\Gamma$ is the Christoffel symbol for flat space in cylindrical coordinates:

\begin{eqnarray}
\Gamma^r_{\phi \phi} = -r \\
\Gamma^{\phi}_{\phi r} = \Gamma^{\phi}_{r \phi} = 1/r,
\end{eqnarray}
and all other components evaluate to zero.  Notation will be used which removes the confusion with indices, introducing the nonindexed $v$ for radial velocity and $\omega$ for angular velocity:

\begin{eqnarray}
v_r = v^r \equiv v,\\ v^{\phi} = \omega, v_{\phi} = r^2 \omega
\end{eqnarray}

Also, for brevity, partial derivatives in radius are replaced with primes:

\begin{eqnarray}
\sigma_{\phi \phi} = 2 \partial_{\phi} ( r^2 \omega ) + 2 r v \\
\sigma_{\phi r} = \partial_{\phi} v + r^2 \omega' \\
\sigma_{r r} = 2 v'
\end{eqnarray}

Note that these expressions for the viscous stress agree with others in the literature, except that these formulas are expressed in a coordinate basis, whereas often an orthonormal basis is used:

\begin{eqnarray}
\sigma_{\hat \phi \hat \phi} = {2 \over r} ( \partial_{\phi} v_{\hat \phi} + v_{\hat r} ) \\
\sigma_{\hat \phi \hat r}    = \partial_r v_{\hat \phi} + {1 \over r} ( \partial_\phi v_{\hat r} - v_{\hat \phi} ) \\
\sigma_{\hat r \hat r}       = 2 \partial_r v_{\hat r}
\end{eqnarray}

This result agrees, for example, with \cite{2006astro.ph..9756E}, equations (10), (11) and (13), except of course that Edgar's result does not set $\eta=0$.

However, this result does not agree at face value with standard expressions in for viscous terms in cylindrical coordinates \citep[e.g.][page 51 of the first edition]{1966hydr.book.....L}.  It is not enough to evaluate the different components of $\tensor \sigma$ in cylindrical coordinates, since there are additional geometric terms which come from taking its divergence.  If $\tensor \sigma$ were a vector, its components would be sufficient, since one can use the divergence theorem to evaluate it.  However, since $\sigma$ is a tensor, additional terms appear.  The appropriate goal, then, is to write $\sigma$ in the following form:

\begin{equation}
(\nabla \cdot \sigma)_j = {1 \over \sqrt{g}}\partial_r ( \sqrt{g} \tilde \sigma^r_{~j} ) + {1 \over \sqrt{g}} \nabla_{\phi} ( \sqrt{g} \tilde \sigma^{\phi}_{~j} ) + S_j
\end{equation}

This expression looks like the divergence of a vector plus a source term, and therefore $\tilde \sigma^i_{~j}$ can be manipulated as a vector (treating the raised index as a vector index).  It is important to point out that the presence of a source term here is inescapable, due to the fact that the geometry does not have a symmetry in the radial direction.

It is straightforward to express the divergence of a tensor in this form, using simple Riemannian geometry:

\begin{equation}
(\nabla \cdot \sigma)_j = {1 \over \sqrt{g}} \partial_i ( \sqrt{g} \sigma^i_{~j} ) - {1 \over 2} \partial_j g_{k l} \sigma^{k l}
\end{equation}

Looking at the radial component:

\begin{equation}
(\nabla \cdot \sigma)_r = {1 \over r} \partial_r( r \sigma^r_{~r} ) + {1 \over r} \partial_{\phi} (r \sigma^{\phi}_{~r} ) - r \sigma^{\phi \phi}
\end{equation}

\begin{eqnarray}
 = {1 \over \sqrt{g}} \partial_r ( \sqrt{g} 2 v' ) \nonumber \\
 + {1 \over \sqrt{g}} \nabla_{\phi} ( \sqrt{g} ( \nabla_{\phi} v + r \omega' - 2 \omega ) )\\
 - 2 v/r^2 \nonumber
\end{eqnarray}

Similarly for the azimuthal component:

\begin{equation}
(\nabla \cdot \sigma)_{\phi} = {1 \over r} \partial_r( r \sigma^r_{~\phi}) + {1 \over r} \partial_{\phi} (r \sigma^{\phi}_{~\phi} )
\end{equation}

\begin{eqnarray}
 = {1 \over \sqrt{g}} \partial_r ( \sqrt{g} ( r^2 \omega' + \partial_{\phi} v ) ) \nonumber \\
 + {1 \over \sqrt{g}} \nabla_{\phi} ( \sqrt{g} ( 2 r^2 \nabla_{\phi} \omega + 2 v ) )
\end{eqnarray}

There are several things to note about these expressions.  First, they also don't agree with standard formulas for viscosity in cylindrical coordinates \citep{1966hydr.book.....L}.  In particular, there are mixed derivative terms.  This can be resolved by choosing a suitable value of $\eta$.  Recall that $\eta$ is the coefficient of the divergence of velocity, which will be abbreviated as $\theta$:

\begin{equation}
\theta = \nabla \cdot v = v' + v/r + \partial_{\phi} \omega
\end{equation}

The $\eta$ term modifies the viscous tress as:

\begin{equation}
\sigma_{i j} \rightarrow \sigma_{i j} + \eta \theta g_{i j}
\end{equation}

Component-wise:

\begin{eqnarray}
\sigma_{r r} \rightarrow \sigma_{r r} + \eta \theta \\
\sigma_{\phi \phi} \rightarrow \sigma_{\phi \phi} + r^2 \eta \theta
\end{eqnarray}

This has the following impact on the divergence:
 
\begin{equation}
\delta (\nabla \cdot \sigma)_r = {1 \over r} \partial_r( r \eta \theta ) - {1 \over r} \eta \theta = \partial_r( \eta \theta )
\end{equation}

\begin{equation}
 = \eta \left( {1 \over r} \partial_r ( r v' ) - { v \over r^2 } + {1 \over r} \partial_{\phi} ( r \omega' )  \right).\\
\end{equation}

Similarly, for the $\phi$ component:

\begin{equation}
\delta (\nabla \cdot \sigma)_{\phi} = \eta \left( {1 \over r} \partial_r( r \partial_{\phi} v ) + \nabla_{\phi} ( r^2 \nabla_{\phi} \omega ) \right)
\end{equation}

There is a natural choice for $\eta$,

\begin{equation}
\eta = -1,
\end{equation}

which simplifies all of these formulas.  This eliminates all of the mixed derivative terms, and restores agreement with standard formulas for viscosity in cylindrical coordinates:

\begin{eqnarray}
(\nabla \cdot \sigma)_r = {1 \over \sqrt{g}} \partial_r ( \sqrt{g} v' ) \nonumber \\
 + {1 \over \sqrt{g}} \nabla_{\phi} ( \sqrt{g} ( \nabla_{\phi} v - 2 \omega ) )\\
 - v/r^2 \nonumber
\end{eqnarray}

\begin{eqnarray}
(\nabla \cdot \sigma)_{\phi} = {1 \over \sqrt{g}} \partial_r ( \sqrt{g} r^2 \omega'  ) \nonumber \\
 + {1 \over \sqrt{g}} \nabla_{\phi} ( \sqrt{g} ( r^2 \nabla_{\phi} \omega + 2 v ) )
\end{eqnarray}

This finally results in the following $\tilde \sigma$:

\begin{eqnarray}
\tilde \sigma^{r}_{~r} = v' \label{eqn:v1} \\
\tilde \sigma^{r}_{~\phi} = \nabla_{\phi} v - 2 \omega \\
\tilde \sigma^{\phi}_{~r} = r^2 \omega' \\
\tilde \sigma^{\phi}_{~\phi} = r^2 \nabla_{\phi} \omega + 2 v
\end{eqnarray}

And a source term for the momentum in the radial direction:

\begin{equation}
S_r = -v/r^2
\label{eqn:v2}
\end{equation}

As expected, there is no source term for angular momentum.  After re-introducing the coefficient $\rho \nu$, equations (\ref{eqn:v1}-\ref{eqn:v2}) can be re-expressed as the terms given in (\ref{eqn:viscflux}) and (\ref{eqn:viscsrc}).

As mentioned in Section \ref{sec:visc_sec}, introducing $\Omega_E(r)$ produces a source term for the energy, given by
\begin{equation}
S^{Energy}_{\rm visc} = \sigma_{r \phi} r \Omega'_E(r).
\end{equation}
using the derived formula for $\sigma$, this can be expressed as

\begin{equation}
S^{Energy}_{\rm visc} = \rho \nu ( \nabla_{\phi} v + r \omega' ) r \Omega'_E(r).
\end{equation}

\end{appendix}


\end{document}